\documentclass[11pt,tightenlines,eqsecnum,floats,aps,amsmath,amssymb,nofootinbib,prd,floatfix]{revtex4}
\usepackage{graphicx, wrapfig}
\usepackage{amssymb}
\usepackage{slashed}
\usepackage{bm}
\usepackage{color}
\usepackage{mathrsfs}
\usepackage{subfigure}
\usepackage{hyperref}

\def\f{\frac}

\def\pp{p_{\phi}}

\def\h{\hat}

\def\hz{\mathring{h}}
\def\ez{\mathring{e}}
\def\wz{\mathring{\omega}}

\def\l{\left}
\def\r{\right}

\def\v{\vec}
\def\A1IJ{{\cal A}_1^{ij}}
\def\A2IJ{{\cal A}_2^{ij}}
\def\A3IJ{{\cal A}_3^{ij}}
\def\A4IJ{{\cal A}_4^{ij}}
\def\A5IJ{{\cal A}_5^{ij}}
\def\A6IJ{{\cal A}_6^{ij}}
\def\A1ij{{\cal A}^1_{ij}}
\def\A2ij{{\cal A}^2_{ij}}
\def\A3ij{{\cal A}^3_{ij}}
\def\A4ij{{\cal A}^4_{ij}}
\def\A5ij{{\cal A}^5_{ij}}
\def\A6ij{{\cal A}^6_{ij}}

\usepackage{enumerate}

\newcommand{\be}{\nopagebreak[3]\begin{equation}}
\newcommand{\ee}{\end{equation}}
\newcommand{\bfig}{\nopagebreak[3]\begin{figure}}
\newcommand{\efig}{\end{figure}}
\newcommand{\bea}{\nopagebreak[3]\begin{eqnarray}}
\newcommand{\ea}{\end{eqnarray}}

\newcommand{\bmult}{\nopagebreak[3]\begin{multline}}
\newcommand{\emult}{\end{multline}}

\usepackage{colordvi}
\usepackage{color}

\definecolor{grenn}{RGB}{0,103,0}

\begin{document}
\title{Observational consequences of Bianchi I spacetimes in loop quantum cosmology}

\author{Ivan Agullo$^{1}$}
\author{Javier Olmedo$^{1,2}$}
\author{V.~Sreenath$^{3}$}
\affiliation{
1. Department of Physics and Astronomy, Louisiana State University, Baton Rouge, Louisiana 70803, USA\\ 
2. Departamento de F\'isica Te\'orica y del Cosmos, Universidad de Granada,  Granada-18071, Spain\\
3. Department of Physics, National Institute of Technology Karnataka, Surathkal, Mangalore 575025, India.}


\begin{abstract}

Anisotropies generically dominate the earliest stages of expansion of a homogeneous universe. They are particularly relevant in bouncing models, since  shears grow in the contracting phase of the cosmos, making the isotropic situation unstable.  This paper extends the study of cosmological perturbations in loop quantum cosmology (LQC) to anisotropic Bianchi I models that contain a bounce followed by a  phase of slow-roll inflation. We show that, although the shear tensor  dilutes and the universe isotropizes soon after the bounce, cosmic perturbations retain memory of this short anisotropic phase. We develop  the formalism needed to describe perturbations in anisotropic, effective  LQC, and apply it to make predictions for the cosmic microwave background (CMB), while  respecting current observational constraints. We show that the  anisotropic bounce induces: (i) anisotropic features in all angular correlation functions in the CMB, and in particular a quadrupolar modulation that can account for a similar feature observed in the temperature map  by the Planck satellite, and (ii) quantum entanglement between scalar and tensor modes, that manifests itself in  temperature-polarization (T-B and E-B)  correlations in the CMB.

\end{abstract}
\maketitle
\section{Introduction}

\label{sec:introduction}

The inflationary scenario \cite{inf,inf1a,inf1b,inf1c,inf1d,inf2,inf2a,inf2b,inf2c,inf2d}  is one of the most promising candidates  to account for the origin of the cosmic structures. It enjoys wide support among cosmologists (see \cite{Steinhardetal} for a different viewpoint). But there is also agreement about the fact that this paradigm is incomplete. Among other things, the big bang singularity of general relativity persists \cite{bgv}. Consequently, it is unclear the way inflation begins, and what sets the initial state for both the background spacetime and the cosmological perturbations. In computing the predictions of inflation, it is common to replace our ignorance about the pre-inflationary universe by suitable ``initial'' conditions. Namely, the background geometry of the observable patch of the universe is assumed to be homogeneous and isotropic. For perturbations, it is common to choose the so-called Bunch-Davies vacuum. These are, however, strong assumptions, supported mainly by the agreement between predictions and current observations. Therefore, this scenario offers an opportunity for models of quantum gravity, since they could shed light on the ultraviolet completion of the inflationary paradigm, and open an observational window into physics beyond inflation. This has been the viewpoint taken in  loop quantum cosmology (LQC) \cite{asrev,lqc,agullo-corichi,lqc2}, where the classical singularity is replaced by a quantum bounce. In this scenario, one can study  the way inflation begins, and how a state close to the Bunch-Davies vacuum for  scalar and tensor perturbations emerges from the  Planck era. But  a limitation remains: in previous studies one begins with a universe that is already homogeneous and isotropic. The goal of this paper is to go a step beyond, by dropping the assumption of isotropy. 

More precisely, this paper is devoted to study, in the context of LQC,  how cosmological perturbations  interact with the anisotropies of the background spacetime, and to investigate under what conditions  primordial anisotropies can  leave  observable imprints in the CMB. These are important questions to be addressed in any bouncing model, since anisotropies grow in the contracting phase before the bounce, and tend to dominate the dynamics as the universe collapses. In LQC, quantum  effects grow faster than any other contribution to the gravitational field equations close to the Planck scale, and  preclude anisotropies from growing unbounded and  reaching a Belinskii-Khalatnikov-Lifshitz type of instability \cite{BKL}. Nevertheless, anisotropies are likely to be significant soon before and after the bounce. But despite the relevance of this subject, the complexity of the calculations has kept researchers away from analyzing anisotropic perturbations in LQC---as well as in other bouncing models---in full detail.

The motivation to embark ourselves in this complex analysis is the following. Gupt and Singh \cite{bi-inf} provided a complete and careful  study of the phenomenology of anisotropic models  of Bianchi I type within the  effective theory of  LQC.  They showed that, on the one hand, the attractor character of inflation remains in LQC and, on the other hand, that by starting from arbitrary anisotropic initial conditions the universe quickly isotropizes either before or soon after the beginning of inflation. This indicates that the consequences of the so-called ``cosmic no-hair theorem'' of general relativity \cite{wald83}, by which anisotropies in the early universe are generically washed out, remains  true if the big bang singularity is replaced by the bounce of LQC.  One of the main messages of our analysis is that perturbations retain memory of the anisotropies for {\em much longer} than the  background geometry does. This is because, while the shear tensor of the Bianchi I metric is proportional to the inverse cube  of the mean scale factor $a$,\footnote{In the absence of anisotropic sources, i.e.\ of matter fields with nonzero anisotropic stresses.}  anisotropies in quantum perturbations only redshift \cite{agulloparker}, and redshift scales inversely  with $a$ (rather than its cube).  Consequently,  unless the inflationary phase is significantly longer than the minimum amount required, perturbations can evade the ``cosmic no-hair theorem'' and  leave an imprint in the cosmic microwave background (CMB). 

Loop quantum cosmology provides an important  advantage compared to general relativistic models. As it was emphasized in \cite{ppu-BI2,pu-BI}, in big bang anisotropic scenarios there is no universal way of choosing an initial state for perturbations, and this  jeopardizes the predictive power of these models.  This is because one or two of the directional scale factors can bounce when propagated backwards in time, before reaching the big bang singularity. Consequently, not all wavelengths of perturbations contract to the past and find an adiabatic regime. In the absence of a preferred initial state, the predictions depend on one's choice. The situation is different in the  bouncing spacetimes we consider in this paper, since the spacetime anisotropies dilute before (and after) the bounce, in such a way that all wavelengths of perturbations that we can prove in the CMB find an adiabatic regime if we go far enough into the past. 

Since LQC is formulated in a canonical or Hamiltonian language, to carry out our analysis we  also need a similar  description of  perturbations in Bianchi I spacetimes. The canonical theory of gauge invariant perturbations in Bianchi I has been developed in a companion paper \cite{aos} (see \cite{ppu-BI1} for a previous analysis based instead on an expansion of Einstein's equations). Reference  \cite{aos}  contains also a detailed description of the quantization of these perturbations. We apply this formalism here to  evolve the quantum perturbations across the bounce, and until  they impact the CMB. We show that an important  difference with the isotropic case is that the presence of anisotropies {\em couples} scalar and tensor perturbations among themselves, and  these couplings induce entanglement in the quantum theory. Hence, the anisotropic bounce produces a quantum state for perturbations that at the beginning of inflation contains both anisotropic features and entanglement between different types of perturbations. The latter translates into nontrivial primordial cross-correlations, that  vanish in isotropic scenarios. We then use these results  to compute the angular correlation functions in the CMB  for temperature and polarization,  contrast the results with existing observations, and make concrete predictions that can be used in the future to test our ideas. 

The rest of the paper is organized as follows. In Sec. \ref{sec:class} we summarize the classical theory of Bianchi I spacetimes, including a brief reminder of the definition of Ashtekar variables, and describe gauge invariant perturbation propagating thereon. Section \ref{sec:quant} deals with the quantum theory. It  first  summarizes the effective theory of  Bianchi I geometries in LQC and its phenomenology, and then describes the quantization of perturbations. Section \ref{sec:pred} is devoted to computing the primordial power spectra of scalar and tensor perturbations, including cross-correlations, and to constrain the parameters of our model  by using observations of the CMB. In Sec. \ref{angularps} we use the primordial power spectra to compute the angular correlation functions in the CMB. We finish in Sec. \ref{sec:discuss} by discussing our results, and by adding some concluding remarks. Appendices \ref{app:lqc} and \ref{app:swsh} contain some details and calculations that have been omitted in the main body of this article.

\section{Classical theory}\label{sec:class}

We summarize in this section the classical theory of Bianchi I spacetimes, minimally coupled to a scalar field with potential $V(\phi)$, together with linear perturbations propagating thereon. We adopt a Hamiltonian formulation \`a la Arnowitt-Deser-Misner (ADM) \cite{Arnowitt:1962hi}, since this is the framework in which the quantum theory, described in the next section, is formulated. The content of this section is well known, therefore we only provide a short summary to make this paper self-contained. Readers can find further details in the original references cited below.

\subsection{Bianchi I spacetimes in general relativity} \label{sec:class-back}

 \subsubsection{Metric variables}
 Consider a three dimensional  manifold $\Sigma$ with $\mathbb{R}^3$ topology. It will be useful for the calculations below to define a (nonphysical) auxiliary flat Euclidean metric $\hz_{ab}$ in $\Sigma$, with line element $d\mathring{s}^2=dx_1^2+dx_2^2+dx_3^2$, where $x_i$ are Cartesian coordinates. Let us also consider an auxiliary box of finite volume $\mathcal{V}_0$ (defined with respect to $\hz_{ab}$).\footnote{\label{V0}From the point of view of general relativity, the integrals on $\Sigma$, like the one involved in the symplectic form, diverge when restricted  to Bianchi I spacetimes,  because of the homogeneity of these geometries. One can avoid this spurious divergence by restricting the integrals to a box,  with arbitrarily large but finite volume  $\mathcal{V}_0$---defined with respect to the auxiliary metric $\hz_{ab}$. We  choose for convenience the sides of the box   aligned with the three axis $x_i$, and of length $L_i$, so  $\mathcal{V}_0=L_1L_2L_3$. One can understand the introduction  of this box  as an infrared regulator, which does not affect physical predictions, and can be taken to infinity at the end of the calculation.} In the ADM formulation of  Bianchi I spacetimes  (see e.g.\ \cite{aos}, and references therein), the phase space is made of four pairs of canonically conjugated variables, ${\bf V}=\{a_i,\pi_{a_i};\phi,p_{\phi}\}$, with $i=1,2,3$. The first three pairs describe the gravitational sector, where $a_i$ are  the ``directional'' scale factors, in terms of which the physical spatial metric takes the form $h_{ab}={\rm diag}(a_1^2,a_2^2,a_3^2)$ in Cartesian  coordinates.\footnote{Bianchi I metrics can always be written in diagonal form when the matter content is a perfect fluid \cite{hawkingellis}. We have identified the Cartesian axes $x_i$ with the  frame in which the Bianchi I metric is diagonal.} Matter is assumed to be made of a real scalar field, with an energy-momentum tensor that has the form of a perfect fluid with energy density $\rho=\dot \phi^2/2+V(\phi)$, pressure $P= \dot \phi^2/2-V(\phi)$, and vanishing anisotropic stresses. $V(\phi)$ is a potential that will be specified later. The nonzero canonical Poisson brackets are 
\be \label{bpb}  \{\phi, \pp\}=\frac{1}{\mathcal{V}_0}\, ,  \hspace{0.5cm} \{a_i, \pi_{a_j} \}=\frac{1}{\mathcal{V}_0} \, \delta_{ij}\, .\ee 
The  degrees of freedom in the phase space $\bf V$ are subject to  one constraint, which originates from the scalar constraint of general relativity---the vector constraints are identically zero when restricted to  Bianchi I geometries, as a result of the homogeneity. The scalar constraint takes the form 
\bea\label{Fcons}  \mathcal{H}_{\rm BI} &=& N\, \f{\mathcal{V}_0}{2\sqrt{h}} \biggl[ 
 \kappa \l( \f{a_1^2 \pi_{a_1}^2}{2} + \f{a_2^2 \pi_{a_2}^2}{2}  + \f{a_3^2 \pi_{a_3}^2}{2} 
 - a_1 \pi_{a_1} a_2 \pi_{a_2} - a_2 \pi_{a_2} a_3 \pi_{a_3} - a_3 \pi_{a_3} a_1 \pi_{a_1} \r)\nonumber\\
 && + p_{\phi}^2 + 2 h \, V(\phi) \biggr] \approx 0,
\ea
where  $h=(a_1a_2a_3)^2$ is the determinant of $h_{ab}$, $\kappa=8\pi G$ with $G$ Newton's constant, and  $N$ is the lapse function: $N=1$ corresponds to cosmic time $t$, while $N=a$ to conformal time $\eta$. This constraint is also the Hamiltonian that generates time evolution by means of  
\bea \label{eoma} \dot a_i&=&\{a_i,  \mathcal{H}_{\rm BI} \} ,\hspace{0.5cm} \dot \pi_{a_i}=\{\pi_{a_i},  \mathcal{H}_{\rm BI}\} \, , \\ \nonumber 
\label{eomphi} \dot \phi&=&\{\phi,  \mathcal{H}_{\rm BI}\} \, , \hspace{1cm} \dot{\pp}=\{\pp, \mathcal{H}_{\rm BI} \} \, .\ea
These  ordinary differential equations are equivalent to Einstein's equations restricted to Bianchi I spacetimes, and contain all information about the dynamics of the coupled system matter-spacetime geometry. We describe now the ``initial'' data that is required to single out a unique physical solution to these equations. On the one hand, the canonical degrees of freedom are not all independent due to the constraint (\ref{Fcons}). On the other hand,  the directional scale factors $a_i$ are not physical observables, since their values  change under a  rescaling of the coordinates $x_i$; only ratios $a_i(t)/a_i(t')$ have intrinsic physical meaning. 
Therefore, a  solution  is uniquely characterized by the value  of $\pi_{a_1}(t_0)$, $\pi_{a_2}(t_0)$, $\pi_{a_3}(t_0)$, $\phi(t_0)$, and the sign of  $\pp(t_0)$ at some instant $t_0$. The values of $a_i(t_0)$ can be chosen arbitrarily without modifying the physical content of the  solution. 

Given a solution $a_i(t),\pi_{a_i}(t),\phi(t),\pp(t)$, one has a complete description of the system. However, it is useful to separate the information contained in these degrees of freedom in aspects that concern only the evolution of any physical volume element (and that ignore anisotropies), and those aspects that are associated to pure anisotropies. The former are given by $\phi$, $\pp$ for the matter sector, and the mean scale factor $a$ and its Hubble rate  $H\equiv \dot a/a$ for the geometry.  They are determined by the canonical variables $a_i$ and $\pi_{a_i}$ by
\be a= (a_1a_2a_3)^{1/3}\, , \ \ {\rm and} \ H=-\frac{\kappa}{6\, a^3}\, \sum_i a_i\pi_{a_i}\, .\ee 
On the other hand, anisotropies are commonly characterized by the anisotropic shears
\be \sigma_i\equiv H_i-H=\frac{\kappa}{a^3} \,a_i\pi_{a_i} +2 H\, , \ \ ({\rm no \ sum \ in }\  i)\, , \ee
where $H_i\equiv \dot a_i/a_i$ (no sum in $i$) are the directional Hubble rates. 
One can check that $H=\frac{1}{3} \sum_i H_i$, which implies that the  $\sigma_i$'s satisfy  $\sigma_1+ \sigma_2+ \sigma_3=0$, so only two of them are independent. The square of the total shear is defined as $\sigma^2=\sum_i \sigma^2_i$, and it is a measure of the degree of anisotropy of a Bianchi I solution, while the $\sigma_i$'s in addition indicate the way these anisotropies are distributed among the three principal directions $ x_i$. 

The physics of  Bianchi I spacetimes in general relativity has been extensively studied in the literature (see e.g. \cite{ppu-BI2} for a recent summary). We summarize here some aspects that are relevant for our analysis. In particular, we restrict to  potentials $V(\phi)$ for the scalar field that are able to produce an inflationary phase.
Under these circumstances,  it has been shown that a phase of slow-roll  is an attractor  of dynamical trajectories  that start from quite arbitrary anisotropic conditions in the far past  (see  \cite{ppu-BI2,bi-inf} and reference therein). This is a consequence of the fact that  in {\em all} solutions to the equations of motion the anisotropic shears $\sigma_i$ are  proportional to $a(t)^{-3}$, which in turn implies that the shear squared  $\sigma^2$ falls off with the expansion  exactly as $a(t)^{-6}$---in other words, the combinations $a(t)^6 \sigma^2(t)$ and $a(t)^3 \sigma_1(t)$ are two independent constants of motion.  As a consequence, the  relevance of the anisotropies decreases with time  relative to the potential energy $V(\phi)$,  and the latter eventually  dominates and brings the universe to an inflationary phase that, furthermore, dilutes anisotropies exponentially fast. A precise formulation of this statement is made in the so-called cosmic no-hair theorem \cite{wald83}.  The details about the way inflation emerges from an earlier anisotropic phase, and the impact of anisotropies on the duration and other aspects of inflation have been studied in  detail in \cite{ppu-BI2,bi-inf}.

On the other hand, Bianchi I geometries  are generically past incomplete, in the  sense that they find  a big bang singularity in a finite amount of proper time in the past. The presence of Weyl curvature makes the  singularity significantly richer than in the isotropic case. In particular, directional scale factors $a_i$ can individually bounce when propagated back in time; one or two of the three $a_i$'s can grow toward the past, while the mean scale factor $a$ tends to zero and the mean Hubble rate $H$ and some curvature invariants diverge. This gives rise to a family of different types of possible  singularities known as point-, cigar-, barrel-, and pancakelike singularities \cite{Thorne} (see also \cite{gs} for a recent analysis).
 
 \subsubsection{Ashtekar variables}

We briefly  summarize  here the evolution of Bianchi I geometries in Ashtekar variables \cite{ashtekarvar1,ashtekarvar2}. With respect to the previous subsection, this is merely a change of variables in the classical theory that does not modify the physics. But, since these are the variables that are used in the quantization approach of loop quantum cosmology discussed below, it is convenient to introduce them  in the simplest context of general relativity. See e.g.\ \cite{bi-inf,bi,bi1,awe} for further details. 
 
Rather than using the spatial metric $h_{ab}$ and its conjugate momentum $\pi^{ab}$ as coordinates in the gravitational sector of the phase space, Ashtekar variables consist of a connection $A^i_a$ and its conjugate variable $E^a_i$ (a densitized triad). These are the analog for gravity of the canonical variables used in Yang-Mills theories; as before, $a$ is a spatial index, and $i=1,2,3$ is a new (internal) index that takes values in the  algebra $su(2)$, and accounts for the $SU(2)$ gauge symmetry that these variables introduce---it is related to the ambiguity in the choice of a triad of orthonormal vectors in space. $E^a_i$ encodes the information of the  metric, $\sqrt{h}\, h^{ab}=E^a_i E^{ib}$, and  $A^i_a$ of the conjugate momenta $\pi^{ab}$.  When restricted to Bianchi I spacetimes, one can fix both the spatial and $SU(2)$ gauge freedoms to write Ashtekar variables in a ``diagonal'' form, in which the information in $A^i_a$ and $E^a_i$ is codified in three numbers $c_i$ and  $p_i$, respectively:
\begin{equation}
 A^i_a\, =\, c_i\,L_i^{-1}\mathring\omega^i_a\,\,\,\, {\rm and}\,\,\,\, 
 E^a_i\, =\, p_i\,L_i\,{\cal V}_0^{-1}\sqrt{\mathring{h}}\,\mathring{e}^a_i,  \ \ \ ({\rm no \ sum \ in}\ i)\, ,
\end{equation}
where $\ez^a_i$ are  three orthonormal vectors with respect to the auxiliary metric $\hz_{ab}$, that point in the direction of the cartesian axes $x_i$; $\wz^i_a$ are the associated covectors, and $L_i$ denote the lengths in each of the principal directions of the auxiliary cell of volume ${\cal V}_0=L_1L_2L_3$ (see footnote \ref{V0}). Therefore, in Ashtekar variables, the  coordinates in the phase space of  Bianchi I geometries are  $c_i$ and $p_i$, together with $\phi$ and $p_{\phi}$. The nonvanishing Poisson brackets are
\be
\{c_i,p_j\}=\kappa \gamma \delta_{ij}, \quad \left\{\phi, p_{\phi}\right\}=\frac{1}{\mathcal{V}_{0}}\, ,
\ee
where $\gamma$ is a new fundamental constant that does not affect physical predictions in the classical theory---although it does after quantization---known as the Barbero-Immirzi parameter \cite{BI}. Its value is suggested by studies  of black hole entropy \cite{meissner,bhcounting,bhcounting2} and is $\gamma=0.237$.
The relation between $c_i$ and $p_i$ and metric variables $a_i$ and $\pi_{a_i}$ is given by\footnote{We restrict here to  $p_i \geq 0$. Negative values of $p_i$ describe the same physics, because the reflections $p_i \to -p_i$ are large gauge transformations.  See \cite{awe,BD} for further details.}
\be p_i=\frac{\mathcal{V}_0}{L_i}\, \frac{a^3}{a_i}\, , \ \ \ \ {c_i=-\kappa \, \gamma \,L_i\, a^{-3} a_i \, \Big(a_i\, \pi_{a_i}-\frac{1}{2}\sum_j a_j\pi_{a_j} \Big)}\, , \ \ \ \ ({\rm no\ sum \ in \ }i)\,  \ee
The Hamiltonian constraint (\ref{Fcons}), when written in terms of $c_i$ and $p_i$, takes the form, 
\be\label{calssHAv}
 \mathcal{H}_{\rm BI}\, =\, N\left[\f{-1}{\kappa\, \gamma^2 \,v}\biggl( 
 c_1 c_2p_1 p_2\, +\,c_1 c_3p_1p_3\,+\,c_2 c_3p_2p_3\,\biggr)\, +\, \f{{\cal V}^2_0p_{\phi}^2}{2v}\, + {v\, V(\phi)}\, \right],
\ee
 where $v=\sqrt{p_1p_2p_3}$. A physical solution to the equations of motion 
\begin{align}
  \dot{c}_{i} &=\left\{c_{i}, \mathcal{H}_{\mathrm{BI}}\right\} , \quad \dot{p}_{i}=\left\{p_{i}, \mathcal{H}_{\mathrm{BI}}\right\}  ,\\ \dot{\phi} &=\left\{\phi, \mathcal{H}_{\mathrm{BI}}\right\} , \quad \dot p_{\phi}=\left\{p_{\phi}, \mathcal{H}_{\mathrm{BI}}\right\}, 
\end{align}
is uniquely singled out by specifying $\phi(t_0)$, $c_i(t_0)$ for $i=1,2,3$, and the sign of  $\pp(t_0)$ at some time $t_0$ (as for the directional scale factors $a_i$, different choices of $p_i(t_0)$ produce physically equivalent solutions). 
The information about anisotropies, that in metric variables was neatly encoded in the constants of motion $ \sigma^2 a^6$ and $\sigma_i a_i^3$, can now be codified in the combinations 
\begin{equation}\label{eq:Oi}
  O_1 = c_1p_1-c_2p_2,\quad O_2 = c_2p_2-c_3p_3,\quad  O_3 = c_3p_3-c_1p_1.
\end{equation}
It is straightforward  to check that these quantities are also constants of motion---i.e.\ they Poisson-commute with the Hamiltonian, $\{O_i,\mathcal{H}_{\rm BI}\}=0$---and contain the same information as $\sigma^2$ and $\sigma_i$. In fact,
since $O_1+O_2+O_3=0$, only two of them are independent.

\subsection{Perturbations} \label{sec:class-pert}

The Hamiltonian theory of gauge invariant, linearized perturbations in Bianchi I spacetimes has been worked out in a companion paper \cite{aos}.
Therefore,  we provide here only a short summary. The physical content of linear perturbations in Bianchi I geometries  can be encoded in three pairs of canonically conjugated fields $\Gamma_{\mu}(\vec{k}),\Pi_{\mu}(\vec{k})$, $\mu=0,1,2$ (we will work in Fourier space, so $\vec k$ labels the wavenumber of a mode with spatial dependence $e^{i\, \vec k \cdot \vec x}$). These fields are gauge invariant, in the sense that they do not change under the gauge transformations generated by the linearized constraints of the theory; or equivalently, they Poisson-commute with the linearized scalar and vector constraints. In the isotropic limit,  the fields $\Gamma_{\mu}(\vec{k})$ reduce to the familiar scalar and tensor perturbations. More precisely, in that limit  $\Gamma_0(\vec{k})=\sqrt{4\kappa}\,  \mathcal Q$, where $\mathcal{Q}$ is the so-called Mukhanov-Sasaki variable\footnote{ $\mathcal{Q}$ is related to comoving curvature perturbations $\mathcal{R}$ by $\mathcal{R}=\frac{a}{z}\, \mathcal{Q}$, where $z=-\frac{6}{\kappa}\frac{p_{\phi}}{p_a}=a\, \frac{\dot \phi}{H}$ and  $p_a$ is the canonically conjugate variable to the mean scale factor $a$ (its relation to $\dot a$ is $p_a=-\frac{6}{\kappa}\, a \, \dot a$).}, and $\Gamma_1$ and $\Gamma_2$ reduce to the $+,\times$ polarizations of tensor modes, respectively. 
The most important difference with respect to cosmological perturbations in Friedmann-Lema\^itre-Robertson-Walker (FLRW) spacetimes is that, in presence of anisotropies, the fields $\Gamma_{\mu}(\vec{k})$ are {\em coupled} to each other. This is manifest by looking at their Hamilton's equations  of motion, that can be combined into the following set of second order ordinary differential equations
\be \label{eqginper}\ddot \Gamma_{\mu}(\vec k)+3\, H\, \dot \Gamma_{\mu}(\vec k)+\frac{k^2}{a^2}\,  \Gamma_{\mu}(\vec k)+\frac{1}{a^2}\, \sum_{\mu'=0}^2\, {\cal U}_{\mu\mu'}\, \Gamma_{\mu'}(\vec k)=0\, , \ee
where dots indicate derivative with respect to cosmic time, and $k^2(t)\equiv a^2(t)\, \left(\frac{k_1^2}{a_1^2(t)}+\frac{k_2^2}{a_2^2(t)}+\frac{k_3^2}{a_3^2(t)}\right)$. The fields $\Gamma_{\mu}(\vec k)$ are coupled by the potentials $ {\cal U}_{\mu \mu'}$ for $\mu\neq \mu'$, which depend on the anisotropies. Their explicit form is given in Appendix \ref{app:lqc} (see Ref. \cite{aos} for additional details). In the isotropic limit, the off-diagonal components of $ {\cal U}_{\mu \mu'}$  vanish, the fields decouple, and one recovers the familiar evolution of scalar and tensor modes in FLRW. 

In deriving physical predictions, as we will see below, it will be more convenient to replace the fields $\Gamma_{1}(\vec k)$ and $\Gamma_{2}(\vec k)$
by the complex combinations 
\be
\Gamma_{\pm 2}(\v k)=\frac{1}{\sqrt{2}}\left(\Gamma_1(\v k) \mp i\, \Gamma_2(\v k)\right)\, .\ee
In the isotropic limit $\Gamma_{\pm 2}(\v k)$ describe right- and left-circularly polarized tensor modes respectively. In other words, they have well-defined helicity $\pm 2$. We will work with these fields in the rest of this paper.

\section{Quantum theory} \label{sec:quant}

In this section we discuss the quantum theory of both Bianchi I geometries and gauge invariant perturbations propagating thereon.

\subsection{Bianchi I spacetimes in loop quantum cosmology}\label{sec:quant-pert}

In loop quantum cosmology, the state of the gravitational field describing  a Bianchi I spacetime is  described  by a wave function $\Psi_{\rm BI}$, that satisfies a Wheeler-deWitt-like equation $\hat{\mathcal H}_{\rm BI}\Psi_{\rm BI}=0$, where  $\hat {\mathcal{H}}_{\rm BI}$ is the operator associated with the classical Hamiltonian constraint ${\mathcal H}_{\rm BI}$. This quantum theory has been developed in  \cite{bi,bi1,awe,bi2,djms,mmwe,qrlg}, and summaries can be found in \cite{asrev,lqc,lqc2}. The main physical aspects of these quantum spacetimes are more clearly understood in the so-called effective theory. This is a classical theory with a quantum corrected  Hamiltonian that encodes the nonperturbative effects of loop quantum cosmology. The solutions to  Hamilton's equations obtained from this effective  Hamiltonian approximate very well the evolution of the peak of (sharply peaked) wave functions $\Psi_{\rm BI}(p_1,p_2,p_3,\phi)$. The advantage is that one can extract and understand the new physics in the Planck regime in a simpler manner. Analytical studies \cite{tav08,wer} and numerical simulations \cite{aps,kimera} have shown that, in FLRW, the effective theory indeed approximates extremely well the evolution of  wave functions $\Psi_{\rm FLRW}(p,\phi)$ that are sharply peaked, and can even describe more general states \cite{mop,dms,ag}. In contrast, in Bianchi I and for the prescription given in \cite{awe}, there are no numerical simulations  of the time evolution of wave functions  $\Psi_{\rm BI}(p_1,p_2,p_3,\phi)$ due to the  complexity of the quantum theory.\footnote{Numerical simulations for other prescriptions \cite{bi2,djms} suggest that the effective dynamics is  valid in these scenarios.} Hence,  the validity of the effective theory 
as a good approximation to the evolution of quantum states,  although seems quite reasonable from the physical viewpoint, is an assumption that remains to be explicitly checked in Bianchi I. We will rest on this assumption in the following.

The effective Hamiltonian in Bianchi I spacetimes in LQC is given by 

\begin{eqnarray}
 \mathcal{H}^{\rm eff}_{\rm BI} &=& N \bigg[ {\f{-1}{\kappa\,\gamma^2\,v}}\biggl( 
 \f{\sin(\bar\mu_1\,c_1)}{\bar\mu_1}\f{\sin(\bar\mu_2\,c_2)}{\bar\mu_2}p_1p_2\, +\f{\sin(\bar\mu_1\,c_3)}{\bar\mu_1} \f{\sin(\bar\mu_3\,c_3)}{\bar\mu_3}p_1p_3+\f{\sin(\bar\mu_2\,c_2)}{\bar\mu_2} \f{\sin(\bar\mu_3\,c_3)}{\bar\mu_3}p_2p_3\biggr)\nonumber\\
 &&+ \f{\mathcal{V}_0^2p_{\phi}^2}{2 v}\, + v\, V(\phi)\bigg].
\end{eqnarray}
It is interesting to note that $ \mathcal{H}^{\rm eff}_{\rm BI}$ can be obtained from the classical Hamiltonian (\ref{calssHAv})  through the replacement $c_i \to  \f{\sin(\bar\mu_i\,c_i)}{\bar\mu_i}$, where $\bar\mu_1\equiv \sqrt{\Delta} \,\sqrt{\f{p_1}{p_2\,p_3}}$, and similarly for $\bar \mu_2$ and $\bar\mu_3$, where  $\Delta = 4 \sqrt{3} \pi \gamma \ell_{\rm{Pl}}^2$ is the so-called area gap, the minimum eigenvalue of the area operator in loop quantum gravity. These trigonometric functions  capture in a precise way the leading order quantum effects of the gravitational field in LQC \cite{awe}. An effective quantum Bianchi I spacetime is then determined by the solutions to the equations of motion generated from $\mathcal{H}^{\rm eff}_{\rm BI}$:

\begin{eqnarray}\label{effeqs}
 \dot c_i\, &=&\, \l\{ c_i\,,\, \mathcal{H}^{\rm eff}_{\rm BI} \r\}\,=\, \kappa\,\gamma\, \f{\partial\,\mathcal{H}^{\rm eff}_{\rm BI}}{\partial\,p_i}\, , \\
 \dot p_i\, &=&\, \l\{ p_i\,,\, \mathcal{H}^{\rm eff}_{\rm BI} \r\}\, =\, -\kappa \,\gamma\, \f{\partial\,\mathcal{H}^{\rm eff}_{\rm BI}}{\partial\,c_i}\, ,\\\dot \phi\, &=&\, \l\{ \phi\,,\, \mathcal{H}^{\rm eff}_{\rm BI} \r\}\,=\frac{1}{\mathcal{V}_{0}}\, \f{\partial\,\mathcal{H}^{\rm eff}_{\rm BI}}{\partial p_{\phi}}\, , \\
 \label{effeqs4}
 \dot p_{\phi}\, &=&\, \l\{ p_{\phi}\,,\, \mathcal{H}^{\rm eff}_{\rm BI} \r\}\, =\, -\frac{1}{\mathcal{V}_{0}}\, \f{\partial\,\mathcal{H}^{\rm eff}_{\rm BI}}{\partial \phi}.
\end{eqnarray}

The phenomenology of the solutions to these  equations, when the matter sector is given by a scalar field with a potential $V(\phi)$ and no anisotropic stresses, has been explored in full detail by Gupt and Singh in \cite{bi-inf}, and we refer the reader there for details.  In summary, the quantum-corrected dynamics of Bianchi I spacetimes is indistinguishable from the predictions of general relativity everywhere except when one or more of the curvature invariants approaches the Planck scale. In that regime, deviations from the classical theory grow quickly, dominate over matter and  shear, and avoid the classical singularity. The main picture is similar to the singularity resolution of isotropic FLRW spacetimes in LQC. But for the same reason as the structure of the  classical singularity  is richer  in Bianchi I spacetimes, the physics that replaces the singularity in LQC is also richer. On the one hand, the energy density of the scalar field $\phi$ is bounded above by $\rho_{\rm max}= 0{.}41$ in Planck units (recall that we use  $\gamma=0.237$), but this upper bound can only be reached in the absence of anisotropies, i.e.\ when $\sigma^2=0$. The shear squared $\sigma^2$ is also bounded above  by $\sigma^2_{\rm max}= 11.57$, again in Planck units. All strong curvature singularities are resolved, as long as the matter sector satisfies the null energy conditions. Furthermore, the classical singularity is replaced by a cosmic bounce of the mean scale factor $a$ in all solutions to the effective equations, where the mean  Hubble rate $H$ vanishes. Generically, neither the energy density nor the shear squared attain their maximum values exactly at the time of the bounce. However, when the shear squared reaches $\sigma^2_{\rm max}= 11.57$, the energy density turns out to reach a value close to $\rho_{\rm max}$ (see \cite{gs2} for more details). On the other hand, during the  quantum gravity phase, the shear squared $\sigma^2$ does not evolve  as $a^{-6}(t)$, as it is the case in the classical theory. So neither $\sigma^2 a^{6}$ nor  $\sigma_i a^{3}$ are constants of motion in the quantum theory. However, the combinations $O_i$, $i=1,2,3$, defined  in Eq. (\ref{eq:Oi}) above, are exact constants of motion both in the quantum as well as in  the classical theory---recall that only two of them are independent.

Another interesting aspect  found in \cite{bi-inf} about the effective dynamics, which will be of relevance for this paper, is that the attractor mechanism of inflation persists in the effective phase space of LQC, and therefore generic solutions to the equations of motion find an inflationary phase at some time to the future of bounce, in which the universe quickly isotropizes. 
Indeed, the presence of shear introduces an additional effective frictional force in the evolution of the scalar fields that makes a phase of  slow-roll to start earlier.

In summary, a generic solution to the effective equations of motion (\ref{effeqs}) - (\ref{effeqs4}) is made of two  solutions to Einstein equations, one contracting in the far past and one expanding in the future, joined together by a cosmic bounce of the mean scale factor $a(t)$. The bounce is caused by quantum gravity effects, and  it does not require the introduction of exotic matter violating the energy conditions of general relativity. Furthermore, the bounce is generic in the sense that it takes place in all solutions. Directional scale factors $a_i(t)$ generically bounce at different times. Deviations from the classical theory appear only when a curvature invariant approaches the Planck scale, and in typical solutions this happens only a few Planck seconds around the time of the bounce. In the classical regime, the shear squared is proportional to $a^{-6}(t)$, and consequently  the universe isotropizes both before and after the bounce.\footnote{Recall that this happens because we are considering  matter with no anisotropic stresses.} Anisotropic shears are therefore maximum in the quantum gravity era of the universe. After the bounce, and in presence of an inflationary potential $V(\phi)$, the universe tends to an inflationary phase of potential domination. The ``length'' of inflation depends on the solution. For a detailed analysis, we refer the readers to  \cite{bi-inf}. 

\begin{figure}[h]
{\centering     
\includegraphics[width = 0.49\textwidth]{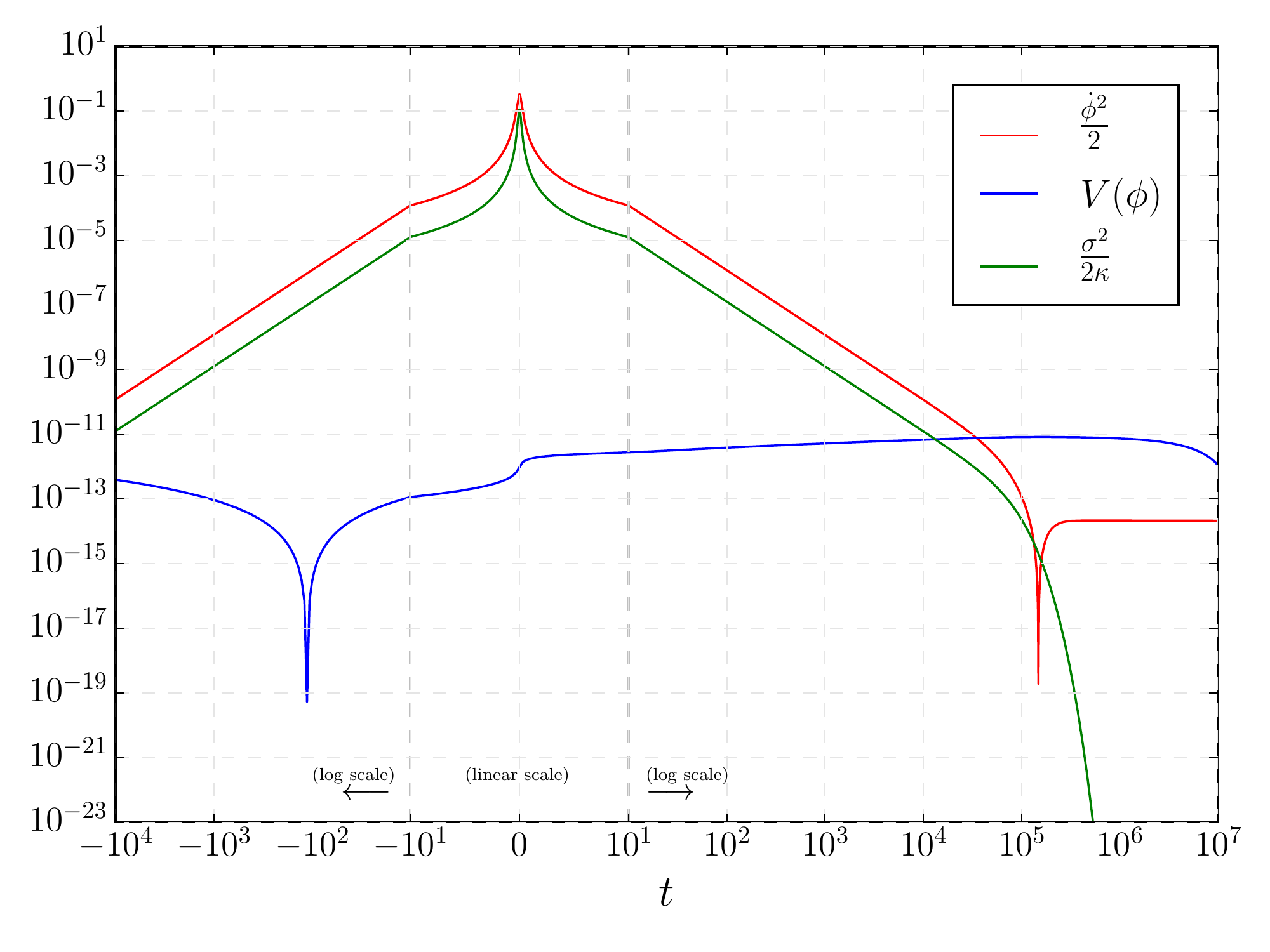}  
  \includegraphics[width = 0.49\textwidth]{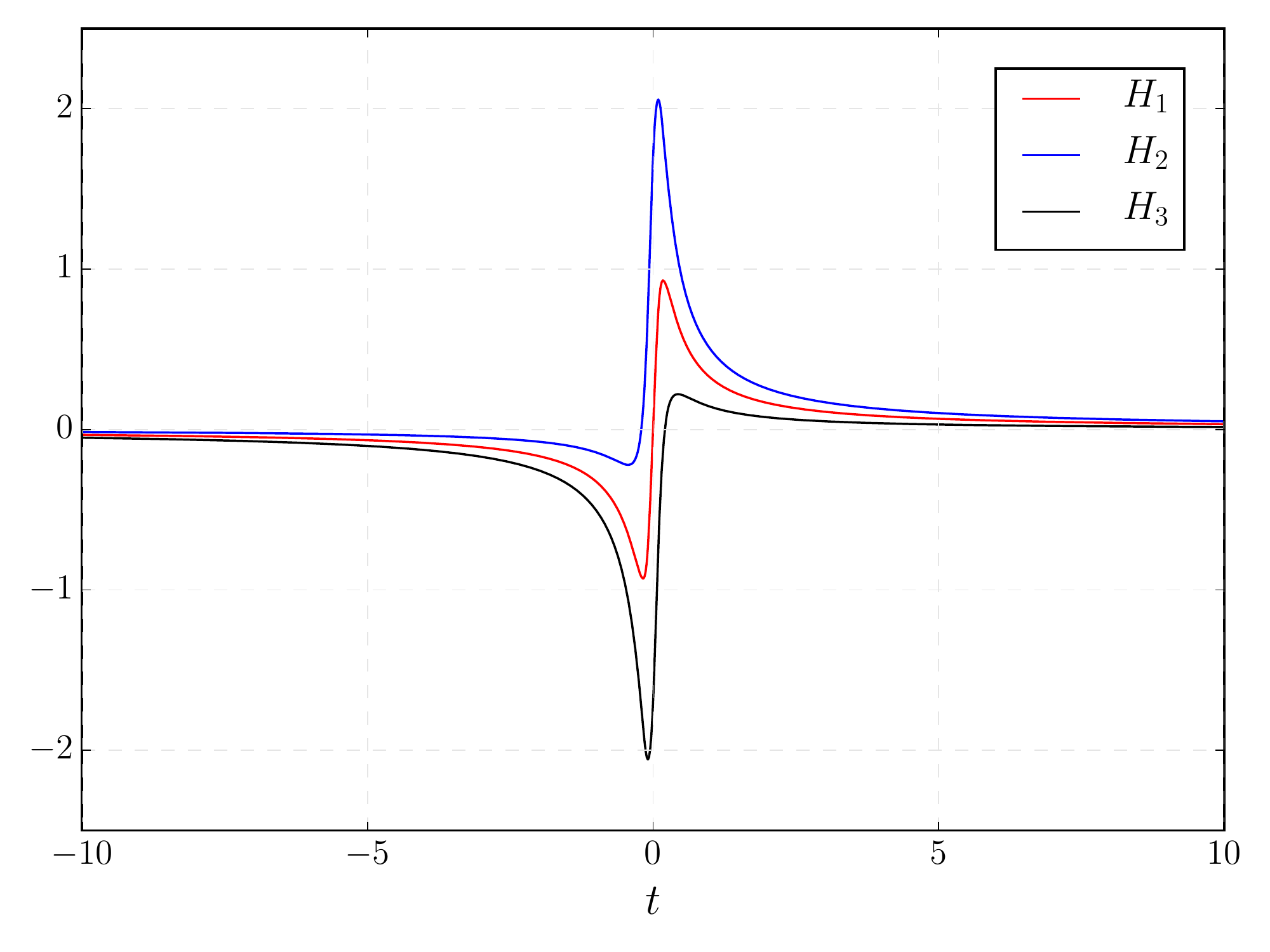}
 }
\caption{ (Left panel)  Kinetic and potential energy densities of the scalar field $\phi$, together with the shear squared (conveniently normalized). The value of $\sigma^2/(2\kappa \rho)$ at the bounce in this simulation is 0.33. This is   half  the value that this ratio would  take if  $\sigma^2(t_B)$ were equal to its supremum value $\sigma^2_{\rm max}= 11.57$. However, notice that, because close to the bounce the general relativistic relation $H^2=\frac{\kappa}{3}\, \rho +\frac{\sigma^2}{6}$ is obviously not satisfied, the ratio $\sigma^2/(2\kappa \rho)$  should not be interpreted as a  measure of the relative contribution of the shear and the energy density of the scalar field to the
 mean expansion rate during the quantum era.\\
(Right panel) Directional Hubble rates around the bounce. This simulation corresponds to $\phi(t_B)=1{.}1$, $\sigma^2(t_B)=5.78$ and $\sigma_1(t_B)=0$, $p_{\phi}(t_B)>0$, and $t_B=0$.}
\label{fig:rho-and-hi}
\end{figure}

We finish this subsection with an example of a typical solution. For the sake of simplicity we choose the quadratic potential $V(\phi)=\frac{1}{2}m^2\phi^2$ for the scalar field, with the mass $m=1.28\times10^{-6}$ in Planck units, obtained from Planck's normalization \cite{plnck2018}. 
Although this  potential is partially disfavored by observations, the physical effects that we describe originate from the bounce, and are largely  independent of  the shape of $V(\phi)$. For illustrative purposes, we show in Fig. \ref{fig:rho-and-hi}  some aspects of the solution to the effective equations for which $\phi(t_B)=1.1$, $\pp(t_B)>0$, $\sigma^2(t_B)=5.78$, and $\sigma_1(t_B)=0$, (this implies  $\sigma_2(t_B)=-\sigma_3(t_B)= \sqrt{\frac{\sigma^2(t_B)}{2}}$), all in Planck units, where $t_B$ is the time at which the mean scale factor bounces. The value chosen for the shear squared is roughly half of its universal upper bound. The left panel shows the evolution of the kinetic and potential energies, together with the shear squared (expressed in units of energy density). The plot shows that while shear and kinetic energy dominate the evolution near the bounce,  anisotropies fall off both in the far past and future, and the universe isotropizes. Around $10^6$ Planck seconds after the bounce, the potential energy dominates the evolution, and a phase of slow-roll inflation begins.  The right panel of Fig. \ref{fig:rho-and-hi} shows the evolution of the directional Hubble rates, together with the mean scale factor. It shows that the three scale factors bounce at different times. For larger values of the shear $\sigma^2(t_B)$ (recall it is bounded by 11.57), the anisotropic phase of the universe extends further to the past and future of the bounce. But it is important to keep in mind that, for the matter content considered in this paper, the universe always isotropizes  away from the bounce \cite{gs,bi-inf}.

\subsection{Perturbations}  \label{sec:quant-back}

The next step is to describe the way the perturbation fields described in Sec. \ref{sec:class-pert} are quantized and how they propagate on the quantum Bianchi I geometries of LQC. The description of  quantum fields propagating on a spacetime that is itself quantum is a challenging problem. But the previous subsection suggests a natural strategy to obtain an approximate solution: quantize the perturbations by treating them as test fields propagating on a smooth spacetime given by the effective geometry of loop quantum cosmology. Indeed, in the isotropic FLRW case  this strategy has been shown to emerge from concrete calculations in LQC. More precisely,  by starting with the simultaneous quantization of both the homogeneous and isotropic degrees of freedom and perturbations, and by neglecting the backreaction of the later in the former, one can derive the equations of motion for the perturbations that, if the quantum spacetime geometry $\Psi_{\rm FLRW}$ is sharply peaked in a classical trajectory at late times, reduce to the propagation of perturbations on the quantum effective geometry. This strategy has been successfully worked out both in the so-called dressed metric approach \cite{aan1,aan2,aan3,am2015}, as well as in the hybrid quantization strategy \cite{hybr-inf1,hybr-inf2,hybr-inf3,hybr-ref,hybr-ten,hybr-pred1,hybr-pred2,hybr-reg}. The two approaches produce similar results, although they differ in the starting point and details. These results have been reinforced by a large number of numerical simulations \cite{dms} that confirm that, even for more general states $\Psi_{\rm FLRW}$ that are not sharply peaked or semiclassical, the peak of the wave function is well described by the effective equations and, therefore, the propagation of perturbations in these effective geometries captures the main effects that the quantum geometry induces in the CMB, at least when the back-reaction can be neglected  (see \cite{ag,aag} for further discussions).

The extension of either the dressed or hybrid quantization strategy for the perturbations to Bianchi spacetimes is straightforward\footnote{The hybrid quantization originally adopted uniqueness criteria for the quantization of perturbations in cosmological spacetimes, as those of Ref. \cite{uniq-perts}. In order to strictly follow this quantization program, these uniqueness results must be extended to cosmological perturbations in Bianchi I spacetimes. Fortunately, they have been understood in the case of a single scalar field \cite{uniq-b1}. We do not see any important obstruction preventing its implementation in the present setting.} and, if the state $\Psi_{\rm BI}$ is assumed to have small quantum dispersions, quantum cosmological perturbations are simply described by the equations of motion (\ref{eqginper}) with the classical Bianchi I metric tensor  replaced by a solution of the effective equations of LQC (\ref{effeqs}) - (\ref{effeqs4}).\footnote{We do not provide here the derivation of the dressed effective metric since, when applied to sharply peaked states $\Psi_{\rm BI}$, it is in fact a trivial extension of the isotropic case \cite{aan1,aan2,aan3,abs}.} We discuss the form of the potentials ${\cal U}_{\mu\mu'}$ in effective LQC in Appendix \ref{app:lqc}. We follow here the strategy that has been successfully used in FLRW, and apply standard techniques of quantum field theory in curved spacetimes---based on Fock quantization---to describe the evolution of perturbations in the Bianchi I effective geometry of LQC. In other words, we neglect potential polymer effects that could affect the physics of quantum fields in full loop quantum gravity. This  is consistent with the level of approximation in the strategy used to quantize the background geometry. We want to emphasize  that, in the absence of a complete theory of quantum gravity, this is the natural strategy that one would follow even in the absence of the theoretical developments spelled out  in \cite{aan1,aan2,aan3,am2015,hybr-inf1,hybr-inf2,hybr-inf3,hybr-ref,hybr-ten,hybr-pred1,hybr-pred2,hybr-reg}. 

The Fock quantization of gauge invariant perturbations in Bianchi I geometries presents an added difficulty with respect to its counterpart in FLRW geometries, namely that scalar and tensor perturbations are coupled to each other, and so we need to quantize an interacting theory. However, as emphasized in \cite{aos}, in spite of these couplings the theory is linear, as it is manifest in equation (\ref{eqginper}), and  this linearity suffices to obtain  an exact, nonperturbative quantization (see \cite{peloso,kioto} for previous analyses). The details of this formulation have been spelled out in detail in the companion paper \cite{aos}, starting from the classical phase space. There, the evolution of perturbations has been formulated  both in the Schr\"odinger and in the Heisenberg pictures, and the way to compute the quantum entanglement between scalar and tensor modes, as well as between the two  tensor modes, induced by the anisotropies of the background geometry, has been shown. Here we provide a short summary of these results in the Heisenberg picture.

The strategy to quantize the coupled system of  scalar and tensor perturbations $\Gamma_s(\vec k)$---which are fields with spin weight $s=0,\pm2$---is similar to the familiar decomposition in normal modes used to solve the dynamics of coupled linear harmonic oscillators. Namely, a representation of the operators $\hat \Gamma_s(\vec k)$ in the Heisenberg picture can be obtained after choosing three independent solutions $\boldsymbol v^{(\lambda)}_{s}(\vec k,t)$ to the coupled system of equations (\ref{eqginper}), labeled by the index $\lambda=1,2,3$, and satisfying the ``Wronskian condition''\footnote{These conditions ensure that the three solutions  $\boldsymbol v^{(\mu)}_{s}(\vec k,t)$ are orthogonal to each other and have positive unit norm with respect to the standard complexified symplectic product. This in turn guarantees that, together with their conjugates, they form a complete basis of the complexified space of solutions to the equations of motion;  furthermore, they define a complex structure in that space, which gives rise to a definition of Fock vacuum. See \cite{aos} for details. On the other hand, in the quantum field theory of perturbations in Bianchi I there are further  ``orthonormality conditions'' that the basis functions need to satisfy, for the map between the algebra of creation and annihilation operators (\ref{fopk}) and the canonical commutation relations of fields and conjugate momenta to be well defined and invertible. Since these extra conditions will not play an important role in this paper, we do not write them here explicitly, and refer the reader to  Sec. IV of \cite{aos} for details.} %
\be\label{Wro}  \sum_{s=0,\pm 2} \bar{ \boldsymbol v}^{(\lambda)}_{s}(\vec k)\, \dot{\boldsymbol v}^{(\lambda')}_{s}(\vec k)-\dot{\bar{ \boldsymbol v}}^{(\lambda)}_{s}(\vec k)\, {\boldsymbol v}^{(\lambda')}_{s}(\vec k)=-i \,\frac{4\kappa}{a^3\, \mathcal{V}_0}\, \delta^{\lambda\lambda'}\, ,\ee
where from now on a bar denotes complex conjugation. If this condition is satisfied at an instant $t_0$, it will hold  at any other time by virtue of the equation of motion. With this condition, the field operators take the form
\be \label{fopk} \hat \Gamma_{s} (\vec k,t)=\sum_{\lambda} \,\left[\boldsymbol v^{(\lambda)}_{s}(\vec k,t)\, \hat a_{\lambda}(\vec k)\, +\bar {\boldsymbol v}^{(\lambda)}_{s}(-\vec k,t)\, \hat a^{\dagger}_{\lambda}(-\vec k)\right]\,, \ee 
where $\hat a_{\lambda}(\vec k)$ and $\hat a^{\dagger}_{\lambda}(\vec k)$  are creation and annihilation operators satisfying 
\be [\hat a_{\lambda}(\vec k),\hat a_{\lambda'}(\vec k')]\,=\,0 \, , \ \ \  [\hat a_{\lambda}(\vec k),\hat a_{\lambda'}^{\dagger}(\vec k')]\,=\,\delta_{\lambda\lambda'}\,\delta_{\vec k, \vec{k}'}\, . \ee
The state annihilated by $\hat a_{\lambda}(\vec k)$ for  $\lambda=1,2,3$ and all values of $\vec k$ is the Fock vacuum, and the action of $\hat a^{\dagger}_{\lambda}(\vec k)$ on it creates excitations associated with the mode ${\boldsymbol v}^{(\lambda)}_{s}(\vec k)$. Hence, ${\boldsymbol v}^{(\lambda)}_{s}(\vec k)$ play the role of the positive frequency modes used in the quantization of fields in flat spacetime. 

\subsubsection{The vacuum state}\label{thevacuum}

As discussed in the introduction and emphasized in \cite{ ppu-BI2}, due to ambiguity in the definition of a vacuum state, general relativistic Bianchi I cosmologies generically lack of predictive power for the cosmic microwave background, unless additional inputs are introduced.
This is because  the strategy based on adiabatic states used e.g.\ in inflation to single out a preferred vacuum, does not always work in anisotropic spacetimes, since in Bianchi I one cannot guarantee that all Fourier modes of interest for the CMB remain in the adiabatic regime in the past.  This is due to the fact that  some of the directional scale factors, when propagated back in time, can bounce and grow to the past, well before reaching the big bang singularity, making the wavelength of Fourier modes that point in their direction to grow to the past and leave the adiabatic regime. In absence of a universal argument to choose the initial state of perturbations, all predictions rest on a choice. 

The situation is different in LQC due to the presence of a bounce of the mean scale factor. Here, the universe always isotropizes in the past for the matter content used in this paper, and therefore perturbations decouple and find an adiabatic regime well before the bounce. We can use that regime to give initial data for perturbations, by selecting an adiabatic vacuum as initial state,\footnote{Strictly speaking, because the adiabatic condition is an asymptotic one in the limit of infinitely large wave numbers, one can still find distinct Fock vacua, all satisfying the adiabatic condition up to some adiabatic order (see e.g.\,  \cite{parker-book}). However, they  typically  produce negligible differences in observable predictions for the power spectrum, and it is for this reason that in the cosmology literature one commonly refers to {\em the} adiabatic vacuum.} recovering in this way the ability to make predictions. We will use $t_0=-10000$ in Planck units (recall that the bounce takes place at $t=0$)  to  prescribe the initial state for perturbations in all the geometries that we will consider in this paper. At that time anisotropies are negligibly small. Because the modes that we can probe in the CMB are well within the adiabatic regime at $t_0=-10000$ and  before, the concrete choice of this initial time does not affect our predictions, i.e.\ one could choose any earlier time and the results would be unaltered.

To prescribe the  vacuum state of perturbations all that we need is to specify a set of  basis functions ${\boldsymbol v}^{(\lambda)}_{s}(\vec k)$ satisfying (\ref{Wro}). Theses modes are uniquely characterized by their value, and the value of their first time derivative at an instant. Our choice for  these quantities at time $t_0=-10000$ is
\bea \label{exp}  {\boldsymbol v}^{(1)}(\vec k)&=&\sqrt{\frac{4\, \kappa }{ a^2\, \mathcal{V}_0}} \frac{1}{\sqrt{2\, k}} \,\left(1,0,0\right) \, , \ \ \  \dot{\boldsymbol v}^{(1)}(\vec k)=\sqrt{\frac{4\, \kappa }{\mathcal{V}_0}}\,\frac{1}{a^2}\,\frac{ -i\, k}{\sqrt{  2\,k}}\, \left(1,0,0\right) \,  , \nonumber \\  {\boldsymbol v}^{(2)}(\vec k)&=&\sqrt{\frac{4\, \kappa }{ a^2\, \mathcal{V}_0}} \frac{1}{\sqrt{2\, k}} \,\left(0,1,0\right)\, , \ \ \  \dot{\boldsymbol v}^{(2)}(\vec k)=\sqrt{\frac{4\, \kappa }{\mathcal{V}_0}}\,\frac{1}{a^2}\,\frac{ -i\, k}{\sqrt{  2\,k}}\, \left(0,1 ,0\right)\nonumber , \\
  {\boldsymbol v}^{(3)}(\vec k)&=&\sqrt{\frac{4\, \kappa }{ a^2\, \mathcal{V}_0}} \frac{1}{\sqrt{2\, k}} \,\left(0,0,1\right)\, ,  \ \ \  \dot{\boldsymbol v}^{(3)}(\vec k)=\sqrt{\frac{4\, \kappa }{\mathcal{V}_0}}\,\frac{1}{a^2}\,\frac{ -i\, k}{\sqrt{  2\,k}}\, \left(0,0,1\right)\, ,\ea
where $k^2\equiv a^{2} k^ik_i$. As we can see, at $t_0$ each of these modes  are excited only  in one of the three perturbations, namely 
$ {\boldsymbol v}^{(1)}(\vec k)$, $ {\boldsymbol v}^{(2)}(\vec k)$ and $ {\boldsymbol v}^{(3)}(\vec k)$ contain excitations only in $\Gamma_0$, $\Gamma_{+2}$ and $\Gamma_{-2}$, respectively (and similarly for the velocities $\dot{\boldsymbol v}^{(\lambda)}(\vec k)$). However, due to the anisotropic couplings between perturbations, at a later time these modes  will generically have nonzero components in all three perturbations.
Using the arguments of \cite{aos}, it is easy to check that the vacuum state defined by these modes  shares the symmetries of the background spacetime at  $t_0=-10000$, namely it is invariant under translations and rotations,\footnote{Strictly speaking, the spacetime is not invariant under rotations since the shear is identically zero only at $t\to -\infty$. However, at $t_0=-10000$, it is small enough to be negligible in the backgrounds we have simulated in this paper.} and also under parity. It is also a vacuum of zeroth adiabatic order.\footnote{Vacua of higher adiabatic order can be defined following  \cite{aan2,ana}. However, as mentioned before, the differences in physical observables would be negligibly small, and  for the sake of simplicity we work with a vacuum of zeroth adiabatic order.} In the Schr\"{o}dinger evolution picture, the state would remain invariant under translations and parity along the entire evolution, because the spacetime itself is also invariant. However,  the  anisotropies in the universe grow  when we approach the bounce, and they  induce anisotropies also in the state of perturbations. These  anisotropies in the perturbations will remain even after the background metric isotropizes in the future, and  can be imprinted  in the CMB. The analysis of these imprints is the goal of the next two sections.

\section{Primordial power spectra} \label{sec:pred}

This section is devoted to the study of the primordial power spectra for scalar and tensor perturbations evaluated at the end of inflation, including cross-correlations between them. We will use these results in the next section to compute the predictions for the angular power spectra in the CMB. This section is organized as follows. We first define the primordial power spectra and describe their properties under rotations and parity transformations; this will help to  understand many features of the angular correlation functions in the next section. We then analyze the results of our model regarding the primordial power spectra, and discuss the physical origin of the new features, which are all related to anisotropies. Finally, we contrast our results with the constraints that the Planck satellite has obtained for anisotropies in the CMB. This will restrict the values of our free parameters; we will use these restrictions in making concrete predictions for the CMB in the next section.

\subsection{Definition and properties}

The primordial power spectra $\mathcal{P}_{ss'}(\vec k)$ are defined from the two-point correlation functions of the fields $\hat \Gamma_{s} (\vec k)$ in Fourier space as

\be \label{psa} \langle 0|\hat \Gamma_{s} (\vec k,t)\hat \Gamma_{s'} (\vec k',t)|0\rangle={\mathcal V}_0^{-1}\ \frac{2\pi^2}{k^3}\, \mathcal{P}_{ss'}(\vec k,t) \, \delta_{\vec k,-\vec k'}\, . \ee
By using the expansion (\ref{fopk}) for the fields  $\hat \Gamma_{s} (\vec k,t)$, one obtains an expression for $\mathcal{P}_{ss'}(\vec k,t)$ in terms of the modes   ${\boldsymbol v}^{(\lambda)}(\vec k,t)$
\be \label{FPS} \mathcal{P}_{ss'}(\vec k,t)={\mathcal V}_0\, \frac{k^3}{2\pi^2} \, \sum_{\lambda}\, \left[{\boldsymbol v}^{(\lambda)}_s(\vec k,t)\,\bar {\boldsymbol v}^{(\lambda)}_{s'}(\vec k,t)\right]\, .\ee
To obtain these spectra  at the end of inflation, all we need to do is to solve the system of second order differential equations (\ref{eqginper}), with initial data given by (\ref{exp}). 

We now enumerate the most relevant properties of $\mathcal{P}_{ss'}(\vec k)$, which will be of great utility in the next subsection (see \cite{aos} for further details).

\begin{enumerate}[(i)]

\item  $\mathcal{P}_{ss'}(\vec k)$ is real and positive definite for $s=s'$, but it is in general  complex if $s \neq s'$. This is obvious from (\ref{FPS}).

\item  $\mathcal{P}_{ss'}(\vec k)= \mathcal{P}_{s's}(-\vec k)$, for all $s$, $s'$. This can be proven from (\ref{psa}) and the fact that field operators $\hat \Gamma_s$ commute among themselves.

\item Under Hermitian conjugation, the fields satisfy $\hat{\Gamma}^{\dagger}_s(\vec k)=\hat{{\Gamma}}_s(-\vec k)$. Consequently, we have that under complex conjugation, $\overline{\mathcal{P}}_{ss'}(\vec k)= \mathcal{P}_{ss'}(-\vec k)$, for all $s$ and $s'$.

\item Under parity, the fields $\hat \Gamma_s(\vec k)$ transform to $\hat \Gamma_{-s}(-\vec k)$; i.e.\ parity interchanges $\hat \Gamma_{+2}$ and  $\hat \Gamma_{-2}$ and inverts the direction of $\vec k$. Therefore, a parity transformation transforms  ${\cal P}_{ss'}(\vec k)$ to  ${\cal P}_{-s-s'}(-\vec k)$.  It is straightforward to check that, for the vacuum state defined by (\ref{exp}), all  spectra  ${\cal P}_{ss'}(\vec k)$ are parity-invariant, i.e.\ ${\cal P}_{ss'}(\vec k)={\cal P}_{-s-s'}(-\vec k)$.  Moreover, together with the  property (ii) this implies that ${\cal P}_{ss'}(\vec k)={\cal P}_{-s'-s}(\vec k)$, and in particular  ${\cal P}_{+2+2}(\vec k)={\cal P}_{-2-2}(\vec k)$.

\item Under rotations, $\hat \Gamma_s(\vec k)$ transform as fields with spin weight $s=0,\pm 2$. Consequently, the power spectra ${\cal P}_{ss'}(\vec k)$ have spin weight $s-s'$. Therefore, to expand  ${\cal P}_{ss'}(\vec k)$ in angular multipoles it is more convenient to  use spin-weighted  spherical harmonics:
\be \label{PssLM} {\cal P}_{ss'}(\vec k)=\sum_{L=|s-s'|}^{\infty}\sum_{M=-L}^L \,   {\cal P}^{LM}_{ss'}(k)\ _{s-s'}Y_{LM}(\hat k) \, , \ee
where $ _{s-s'}Y_{LM}(\hat k)$ is a spherical harmonic of spin weight $s-s'$. This expansion guarantees that ${\cal P}^{LM}_{ss'}(k)$ are scalars under rotations. Note that in the previous expression we have taken into account  that $ _{s-s'}Y_{LM}(\hat k)$ vanish for $L<|s-s'|$. This in turn implies that the isotropic part ($L=0$) of ${\cal P}_{ss'}(\vec k)$  is equal to zero unless $s-s'=0$. Hence  ${\cal P}_{0\pm2}$,   ${\cal P}_{+2-2}$ and ${\cal P}_{-2+2}$ do not have any isotropic mode, and they must vanish  in the isotropic limit. 

On the other hand, the property (iii)  implies that  ${{\cal P}}^{LM}_{ss'}(k)=(-1)^{L+M+s-s'}\, {\bar{\cal P}}^{L-M}_{ss'}(k)$, and this means that in the calculations below it will be sufficient to restrict to $M\geq 0$.

\item Properties (ii) and (iii) above imply that  the real part of ${\mathcal{P}}_{ss'}(\vec k)$ remains invariant under inversion $\vec k \to -\vec k$ (do not confuse inversion  of the wave number  $\vec k$ with a parity transformation, since the later also changes $s\to -s$), while its imaginary part changes sign. On the other hand, since $\mathcal{P}_{ss'}(\vec k)$ is real when $s=s'$ (property (i) above),  the expansion of $\mathcal{P}_{ss}(\vec k)$ contains  only even multipoles $L$.

\end{enumerate}

Finally, in order to compare with observations, it is  useful to report our results involving   scalar perturbations in terms of comoving curvature perturbations $\mathcal{R}$, since this is the variable that is time independent in super-Horizon scales after inflation. 
As  discussed in Sec. \ref{sec:class-pert}, at the end of inflation  $\mathcal{R}$ and  $\Gamma_0$ are related by $\mathcal{R}=\sqrt{4\pi} \frac{\dot \phi}{H}\, \Gamma_0$. Hence, their power spectra are related by\footnote{The power spectra in the basis of tensor modes with linear polarization (typically adopted in isotropic scenarios) are related with the ones in the basis of circular polarization by
\begin{align*}
{\cal P}_{++}(\v k) &= {\cal P}_{22}(\v k)+\frac{1}{2}\left[{\cal P}_{-22}(\v k)+{\cal P}_{2-2}(\v k)\right],\quad {\cal P}_{\times\times}(\v k) = {\cal P}_{22}(\v k)-\frac{1}{2}\left[{\cal P}_{-22}(\v k)+{\cal P}_{2-2}(\v k)\right],\\
{\cal P}_{+\times}(\v k) &= \frac{i}{2}\left[{\cal P}_{2-2}(\v k)-{\cal P}_{-22}(\v k)\right]\,=\,{\cal P}_{\times +}(\v k),\quad {\cal P}_{+{\cal R}}(\v k) =\frac{1}{\sqrt{2}}\left[{\cal P}_{2\mathcal{R}}(\v k)+{\cal P}_{-2\mathcal{R}}(\v k)\right],\\
{\cal P}_{\times{\cal R}}(\v k) &=\frac{i}{\sqrt{2}}\left[{\cal P}_{2\mathcal{R}}(\v k)-{\cal P}_{-2\mathcal{R}}(\v k)\right].
\end{align*}
In isotropic scenarios, the power spectra do not depend on the direction of  $\vec k$, and  ${\cal P}_{++}(k)={\cal P}_{\times\times}(k)={\cal P}_{22}(k)$,  ${\cal P}_{+\times}(k)={\cal P}_{+{\cal R}}(k)={\cal P}_{\times{\cal R}}(k)=0$.
}

\be {\mathcal P}_{\mathcal{R}}(\v k) = \frac{1}{4\kappa} \left(\frac{H}{\dot \phi}\right)^2 \, {\mathcal P}_{00}(\v k) \, , \ \ \ \  {\rm and } \ \ \ \ {\mathcal P}_{\pm 2\mathcal{R}}(\v k) =  \frac{1}{\sqrt{4\kappa}} \left(\frac{H}{\dot \phi}\right) \, {\mathcal P}_{\pm 20}(\v k) \, .\ee

\subsection{Results}

At first sight, we expect that the anisotropic features induced by the spacetime geometry in the perturbations will be larger for wave numbers $\vec k$ with {\em small} ``comoving'' norm  $k$ (defined below Eq.\ (\ref{eqginper})). This is because if $k$ is very large, one can neglect the last term in (\ref{eqginper}) around the time of the bounce, and since this term contains the information about anisotropies, these wave numbers do not ``feel'' the anisotropic bounce. In simpler words, very ultraviolet wave numbers are not affected by the bounce and, consequently, neither by anisotropies. Thus, we expect their primordial power spectrum to be dominated by the physics during the inflationary era, when the universe is already isotropic. In contrast, we expect perturbations with small values of $k$ to be significantly affected by the bounce, and their primordial spectra to have new features relative to the inflationary predictions. More concretely, for infrared modes we expect both anisotropies and deviations from scale invariance. 

As explained in previous sections, the free parameters in our model are coming from the  background spacetime, i.e.\ from the choice of solution to the effective equations (\ref{effeqs}) - (\ref{effeqs4}). The initial data required to single out a unique physical solution of these equations was explained in Sec. \ref{sec:class-back}. It turns out that it is more convenient to specify the initial data at the time $t_B$ of the bounce, since the mean Hubble rate vanishes there, $H(t_B)=0$, and consequently  the values of $\phi(t_B)$, $\sigma^2(t_B)$ and $\sigma_1(t_B)$, together with the sign of $p_{\phi}(t_B)$, suffice to single out a solution.\footnote{Recall that the value of $p_i$, or equivalently of the directional scale factors, do not change the physical solution. For convenience, we choose  $p_i$ by selecting the mean scale factor to be equal to one at the bounce, $a(t_B)=1$, and the there directional scale factors to agree with each other at late times when the universe isotropizes. More concretely, we choose $a_1(t_e)=a_2(t_e)=a_3(t_e)$, where $t_e$ indicates the end of inflation.} We will focus on positive $p_{\phi}(t_B)$ since, as discussed below, these are the values that will produce interesting phenomenology. Hence, the relevant free parameters for us are $\phi(t_B)$, $\sigma^2(t_B)$ and $\sigma_1(t_B)$. The last two measure the amount of anisotropies at the time of the bounce and the way they are distributed among the principal directions, respectively. On the other hand,  the value of $\phi(t_B)$ controls the amount of expansion that the universe accumulates between  the bounce and the end of inflation---the larger  $\phi(t_B)$ is, the larger this expansion is. $\sigma^2(t_B)$ also affects the amount of expansion, but in a smaller amount.

Let us now discuss the results of our numerical simulations for the primordial power spectra. We start by considering a background geometry with $\phi_{B}=1{.}1$, $\sigma^2(t_B)=5.78$ and $\sigma_1(t_B)=0$, all in Planck units. This choice of $\sigma^2(t_B)$ is  half the value of the universal upper bound, $\sigma^2_{\rm max}= 11.57$; i.e., we are considering a universe that is significantly  anisotropic in the  quantum phase. There are $70.1$ $e$-folds of expansion between the bounce and the end of inflation in this solution.  This value agrees with the results found in \cite{barrau} for  the preferred value of $N$ in anisotropic LQC.  We will discuss later the results for other choices. As mentioned at the end of Sec. \ref{sec:quant}, we begin the evolution at $10000$ Planck times before the bounce, where all modes of relevance for the CMB are in the adiabatic regime, and the state of perturbations is an adiabatic vacuum characterized by (\ref{exp}).  We evolve the perturbations across the bounce, until the end of inflation, and compute the value of the multipolar components of the power spectra ${{\cal P}\,}_{ss'}^{LM}(k)$, for $s,s'=\mathcal{R},\pm2$. It will be useful to keep in mind that the range of physical wave numbers that we can directly probe in the CMB  ranges form $10^{-4}{\rm Mpc}^{-1}$ to $10^{-1}{\rm Mpc}^{-1}$. We will plot our power spectra as a function of $(k/k_\star)$, where $k_\star$ is the pivot scale used by Planck \cite{plnck2018},  and whose physical value today is $0.05 {\rm Mpc}^{-1}$. So the observable window corresponds approximately to $(k/k_\star)\in [0.002,4]$. Some details about computational aspects of the  numerical simulations are summarized in Sec. \ref{sec:numerics}.  

\begin{enumerate}
\item {\bf Scalar-Scalar power spectrum.} We report here the results for the multipolar components of the scalar power spectrum, ${{\cal P}}_{\mathcal{R}}^{LM}(k)$. As explained in the previous subsection, ${{\cal P}}_{\mathcal{R}}(\vec k)$ is real and positive, and it only contains even multipoles $L$ due to parity invariance. 

We plot in the left panel of Fig. \ref{fig:P00} the isotropic  multipole $L=0,M=0$, and compare it with an almost scale-invariant power spectrum, as the one predicted by the standard inflationary scenario. For the values of $\phi(t_B)$, $\sigma^2(t_B)$, $\sigma_1(t_B)$ we have chosen in this simulation, this plot shows that ${{\cal P}}^{00}_{\mathcal{R}}(k)$ becomes indistinguishable from  the results of inflation for large values of $(k/k_\star)$, concretely for $(k/k_\star)\geq 0.02$. So, as expected, the effects of the bounce  are restricted to  the most infrared scales in the CMB, and they manifest in the isotropic scalar multipole ${{\cal P}}^{00}_{\mathcal{R}}(k)$ in a deviation from scale invariance. 

The right panel in Fig. \ref{fig:P00} shows the multipoles $L=2$, the first non-zero anisotropic multipoles of the scalar power spectrum. Again, as expected, these multipoles fall off quickly with $k/k_\star$, but they are different from zero.   These anisotropies are a ``memory'' from the anisotropic bounce and, although  they are restricted to the infrared part of the spectrum, we see that they are not necessarily washed out by inflation. As we will show in the next subsection, for some choices of the free parameters of the model, these anisotropies can be so large that they are already  ruled out by observations. 

We have computed  ${{\cal P}}_{\mathcal{R}}^{LM}(k)$ up to $L=7$, since higher multipoles require prohibitively large numerical resources. All multipoles with $L\geq 4$ that we have computed show similar features as  $L=2$, regarding scale dependence and amplitude.

\begin{figure}[h]
{\centering     
\includegraphics[width = 0.49\textwidth]{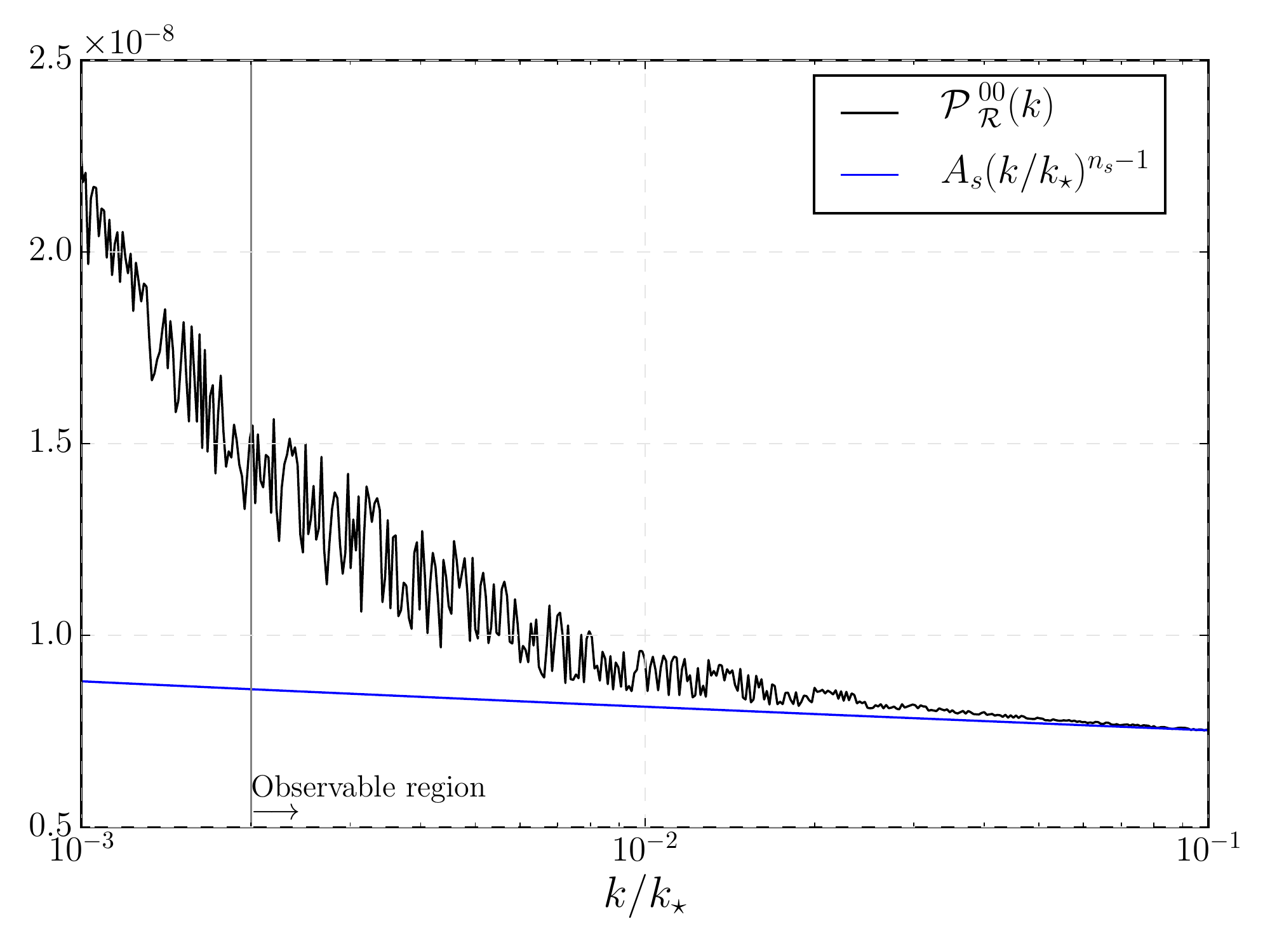}
\includegraphics[width = 0.49\textwidth]{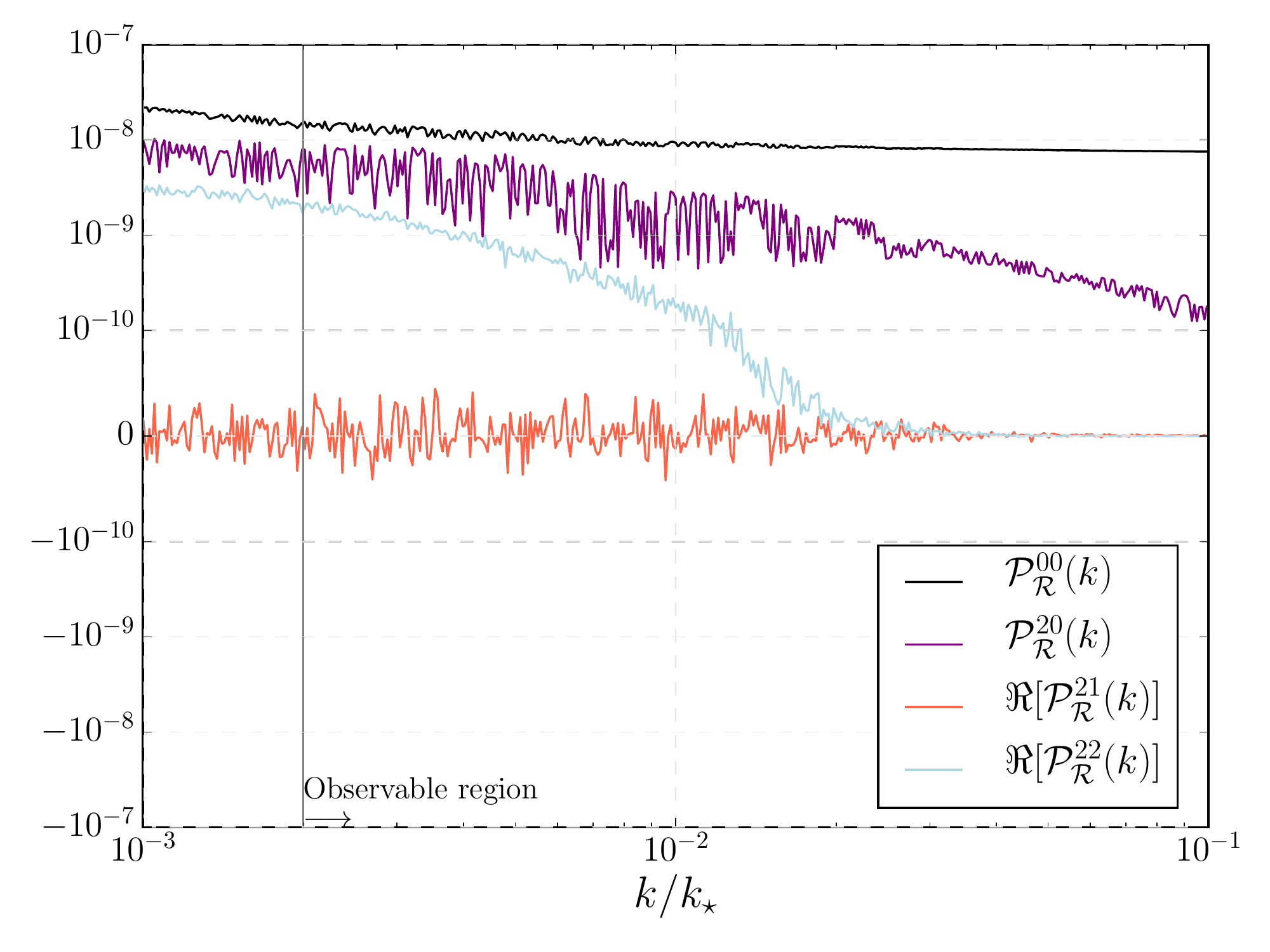}}
\caption{
Multipolar components of the scalar power spectrum ${{\cal P}}_{\mathcal{R}}^{LM}(k)$, for $L=0,M=0$ (left panel) and  $L=2,M=0,1,2$ (right panel)---multipoles with negative $M$ are determined by those with $M$ positive, thanks to the reality condition of ${{\cal P}}_{\cal R}(\v k)$. This simulation corresponds to  $\phi(t_B)=1{.}1$, $\sigma^2(t_B)=5.78$ and $\sigma_1(t_B)=0$.  In the left panel we compare ${{\cal P}}^{00}_{\mathcal{R}}(k)$ with an almost scale-invariant, isotropic, scalar power spectrum $A_s\, (k/k_\star)^{n_s-1}$, with  spectral index $n_s= 0{.}966$. The effects of the bounce are restricted to $(k/k_\star)\leq 0.02$, and they break scale invariance. In the right panel, we show the multipolar components ${{\cal P}}_{\mathcal{R}}^{LM}(k)$, for $L=2$ and $M=0,1,2$ ($L=0$ has been included to help compare their amplitudes). We do not show the imaginary parts of ${{\cal P}}_{\mathcal{R}}^{LM}(k)$ because, for this particular simulation, they turn out to be subdominant with respect to the real ones.\\}
\label{fig:P00}
\end{figure}

\item {\bf  Diagonal Tensor-Tensor power spectrum.} We show in Fig. \ref{fig:P22} the results for the multipolar components of the tensor power spectrum, ${{\cal P}}_{22}^{LM}(k)={{\cal P}}_{-2-2}^{LM}(k)$. As in the scalar  case, they  only contain even multipoles $L$. The features of these spectra are similar to the ones of ${{\cal P}}_{\cal R}^{LM}(k)$, namely the isotropic part $L=0$ shows a deviation from scale invariance for $(k/k_\star)\leq 0.02$ (with a stronger relative enhancement of power than the one of ${{\cal P}}_{\cal R}^{00}(k)$), while it agrees with the inflationary predictions at more ultraviolet scales. The anisotropic multipoles are large also for those infrared scales, and tend to zero for $(k/k_\star)\geq 0.02$.

One interesting  message from these plots is that, in the presence of anisotropies, scalar and tensor perturbations are affected differently by the bounce. This in turn implies that the tensor to scalar ratio $r(\vec k)=({{\cal P}}_{22}(\vec k)+{{\cal P}}_{-2-2}(\vec k))/{{\cal P}}_{\cal R}(\vec k)=2\ {{\cal P}}_{22}(\vec k)/{{\cal P}}_{\cal R}(\vec k)$ is altered with respect to the standard inflationary predictions in the isotropic limit \cite{aan3,hybr-pred2}. In particular, for infrared scales in the CMB it depends on both the norm and the direction of $\vec k$.

\begin{figure}[h]
{\centering     
\includegraphics[width = 0.79\textwidth]{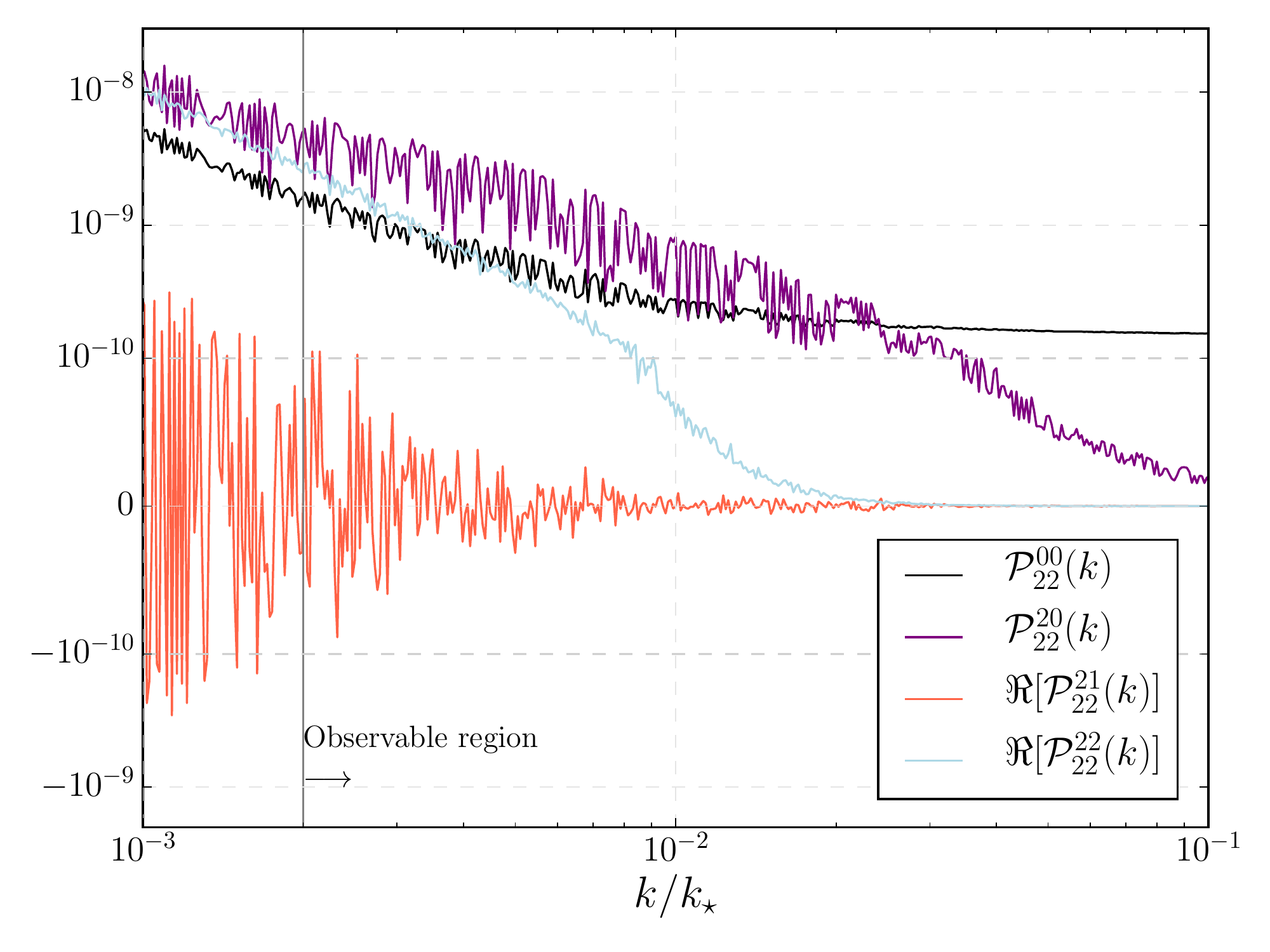}  }
\caption{
Multipolar components ${\cal{P}}_{22}^{LM}(k)$, for $L=0,2$, and $M=0,1,2$. Recall that ${\cal{P}}_{22}^{LM}(k)={\cal{P}}_{-2-2}^{LM}(k)$. The isotropic $L=0$ part tends to an almost scale-invariant spectrum for  large $k/k_{\star}$, while the anisotropic multipoles tend to zero. For this particular simulation, the imaginary parts of ${{\cal P}}_{22}^{LM}(k)$ turn out to be negligible compared to the real ones. Therefore, we do not show them here. Besides,  multipoles with $M<0$ are determined from those with $M$ positive by ${\cal{P}}_{22}^{L-M}(k)=(-1)^M\, \bar{\cal{P}}_{22}^{LM}(k)$.}
\label{fig:P22}
\end{figure}

\item {\bf Scalar-Tensor cross-correlations.} From the properties listed  at the end of the previous subsection, we can see that all cross-correlations between tensor and scalar modes can be determined, for instance, from ${\cal P}_{2\cal{R}}(\v k)$, and therefore we will focus on this quantity. This power spectrum is complex and contains both even and odd multipoles for $L\geq 2$---hence it is purely anisotropic. Figure \ref{fig:P20} shows ${\cal P}_{2\cal{R}}^{LM}(k)$ for $L=2,3,4$ and 5. Their amplitude, although significantly smaller than the diagonal spectra, is different from zero. These correlations are a smoking gun of the anisotropies of the preinflationary universe.
\begin{figure}[h]
{\centering     
\includegraphics[width = 0.79\textwidth]{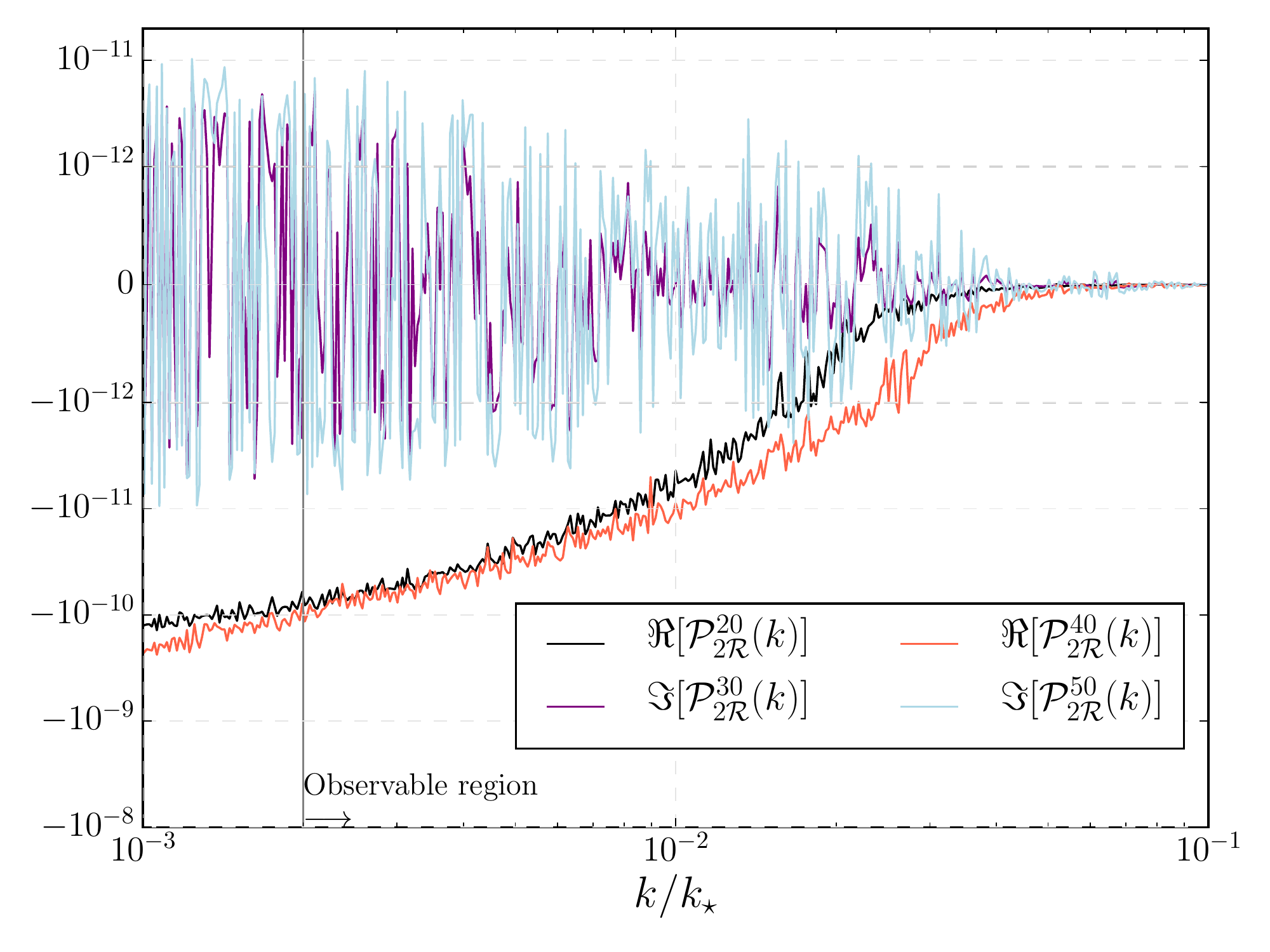}  }
\caption{
Multipolar components of the tensor-scalar spectrum ${{\cal P}\,}_{2\cal{R}}^{LM}(k)$, for $L=2,3,4,5$, and $M=0$. These cross-correlations vanish in an isotropic universe. Multipoles with $M\neq 0$, for a given $L$, show similar features as the ones for $M=0$. }
\label{fig:P20}
\end{figure}

\item {\bf Tensor-Tensor cross-correlations.} Anisotropies also generate cross-correlations between tensor modes, that are described by  ${\cal P}_{-22}(\v k)$ and ${\cal P}_{2-2}(\v k)$, which are also complex---we will focus on the former, since ${\cal P}_{2-2}(\v k)$ can be determined from it using the properties listed above. The spectrum ${\cal P}_{-22}(\v k)$ has spin-weight equal to minus four. Thus, its multipoles are different from zero only for $L\geq 4$. Figure \ref{fig:Pm22} shows some of these multipoles, concretely those that will contribute more to the angular correlation functions discussed in the next section. Interestingly, even multipoles show amplitudes comparable to ${\cal P}^{00}_{22}(\v k)$ at infrared scales , while odd multipoles have smaller amplitudes. This is a manifestation of the fact that the enhancement of power is asymmetric for the $+$ and $\times$ linear polarizations of tensor modes. 
\begin{figure}[h]
{\centering     
\includegraphics[width = 0.79\textwidth]{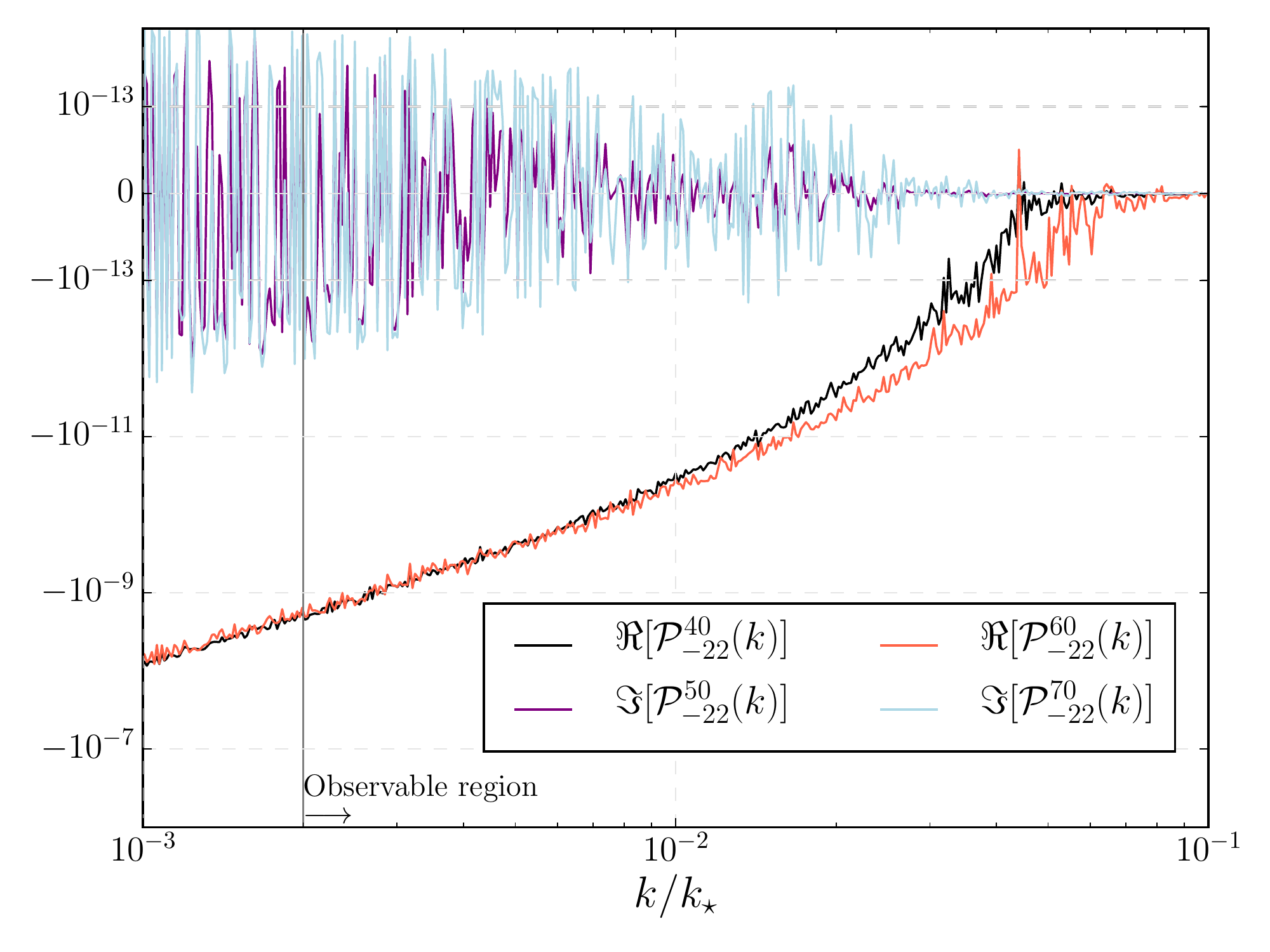}  }
\caption{
Multipolar components of the tensor-tensor cross-correlations ${{\cal P}\,}_{-22}^{LM}(k)$, for $L= 4,5,6,7$, and $M=0$. These correlations vanish in an isotropic universe. We do not show $M\neq 0$ since their behavior is similar to the $M=0$ multipole.} 
\label{fig:Pm22}
\end{figure}

\end{enumerate}

How do these plots depend on the free parameters $\phi(t_B)$, $\sigma^2(t_B)$ and $\sigma_1(t_B)$? On the one hand, since the role of $\sigma_1(t_B)$ is to indicate the way  anisotropies are distributed among the principal directions, a change in $\sigma_1(t_B)$  merely changes the relative size of ${{\cal P}}_{ss'}^{LM}(k)$ for different $M$'s. On the other hand,  $\sigma^2(t_B)$ controls the total amount of anisotropies; the main effect of changing  it is a re-scaling of ${{\cal P}}_{ss'}^{LM}(k)$ for $L>0$, but the dependence on $k$ remains qualitatively the same. Finally, recall that the main role of $\phi(t_B)$ is to control the number of $e$-folds of expansion accumulated after the bounce---larger $\phi(t_B)$ produces more expansion. Hence, by decreasing $\phi(t_B)$  the effects of anisotropies are shifted toward more ultraviolet scales. Similarly, by increasing $\phi(t_B)$  all effects caused by the anisotropic bounce are shifted toward infrared scales, and possibly out of the observable universe for large enough $\phi(t_B)$.

\subsection{Constraints from observations}

CMB observations have revealed some traces of  anisotropies  \cite{plnck2018}. More concretely, the Planck satellite has measured a nonzero amplitude for the leading order deviation from isotropy in a parity-invariant universe, i.e.\ a quadrupole on the scalar power spectrum, ${{\cal P}}_{\cal R}^{LM}$ for $L=2$. However, the statistical significance of this detection is low, compatible with cosmic variance in an isotropic universe.\footnote{On the other hand, both WMAP and Planck satellites have observed a dipolar anisotropy with modest significance of approximately three standard deviations \cite{WMAPdipol,Planck13,Planck15,Planck18}. Furthermore, this dipolar modulation is  observed only at large angular scales. A dipolar modulation in the scalar power spectrum breaks parity invariance, and in consequence it cannot arise in a Bianchi I type universe, unless additional physics that breaks this symmetry is introduced.}  Such a quadrupolar component on  the primordial power spectrum arises naturally in our model. The goal of this section is to derive the implications that  Planck's observations of this quadrupole  have for the free parameters of our model. We will use the results in the next section to work out the predictions of our model.

More concretely, we want to find the values of our free parameters $\phi(t_B)$ and $\sigma^2(t_B)$ that make the CMB as anisotropic as allowed by Planck's observations.\footnote{Although $\sigma_1(t_B)$ is also a free parameter, it is irrelevant for the purpose of this section since, as discussed above, it carries no information about the total amount of anisotropies, but only about the way they are distributed among the principal directions.}  To find these values, we will proceed as follows. We will fix  $\sigma^2(t_B)$ to be close to its maximum value, and will decrease $\phi(t_B)$ until Planck's constraint is saturated. The minimum value of $\phi(t_B)$ compatible with Planck's constraint will produce the most anisotropic CMB allowed by current data. 

Let us first discuss Planck's constraints on the amplitude of the quadrupolar component of the scalar power spectrum (see \cite{plnck2018} for details). The Planck team  considers a phenomenological model of a scalar power spectrum that contains a quadrupolar modulation of the form 
\be \label{Planckpar}
   {\cal P}_{\cal R}(\v k)=\frac{1}{\sqrt{4\pi}}{\cal P}^{00}_{\cal R}(k)\left(1+\sum_{M=-2}^{M=2}g_{2M}(k)\, Y_{2 M}(\hat k)\right)\, ,
\ee
(the factor $1/\sqrt{4\pi}$  comes from the spherical harmonic $Y_{00}=1/\sqrt{4\pi}$) where $g_{2M}(k)$ parameterizes the amplitude of the quadrupole relative to the monopole, and it is allowed to depend on $k$ (hence, this is a scale-dependent quadrupole). The analysis of \cite{plnck2018} only considers  scale dependence of a power law type, of the form  $g_{2M}(k)= g_{2M}\, \left(\frac{k}{k_\star}\right)^q$, with $k_\star=0.05\, {\rm Mpc}^{-1}$  a reference scale, and restricts to $q=0,\pm 1,\pm 2$.  For $q=0$, equation (\ref{Planckpar}) models a scale-independent quadrupole, while for positive (negative) $q$ this is a  blue (red) tilted quadrupolar modulation.  By comparing with CMB data, reference \cite{plnck2018} extracts the mean value of  $g_{2M}$, i.e. $g_2\equiv \sqrt{\sum_M |g_{2M}|^2/5}$, in the CMB, for different choices of $q$  (see Table 17 in \cite{plnck2018}).

\begin{figure}[h]
{\centering     
\includegraphics[width = 0.75\textwidth]{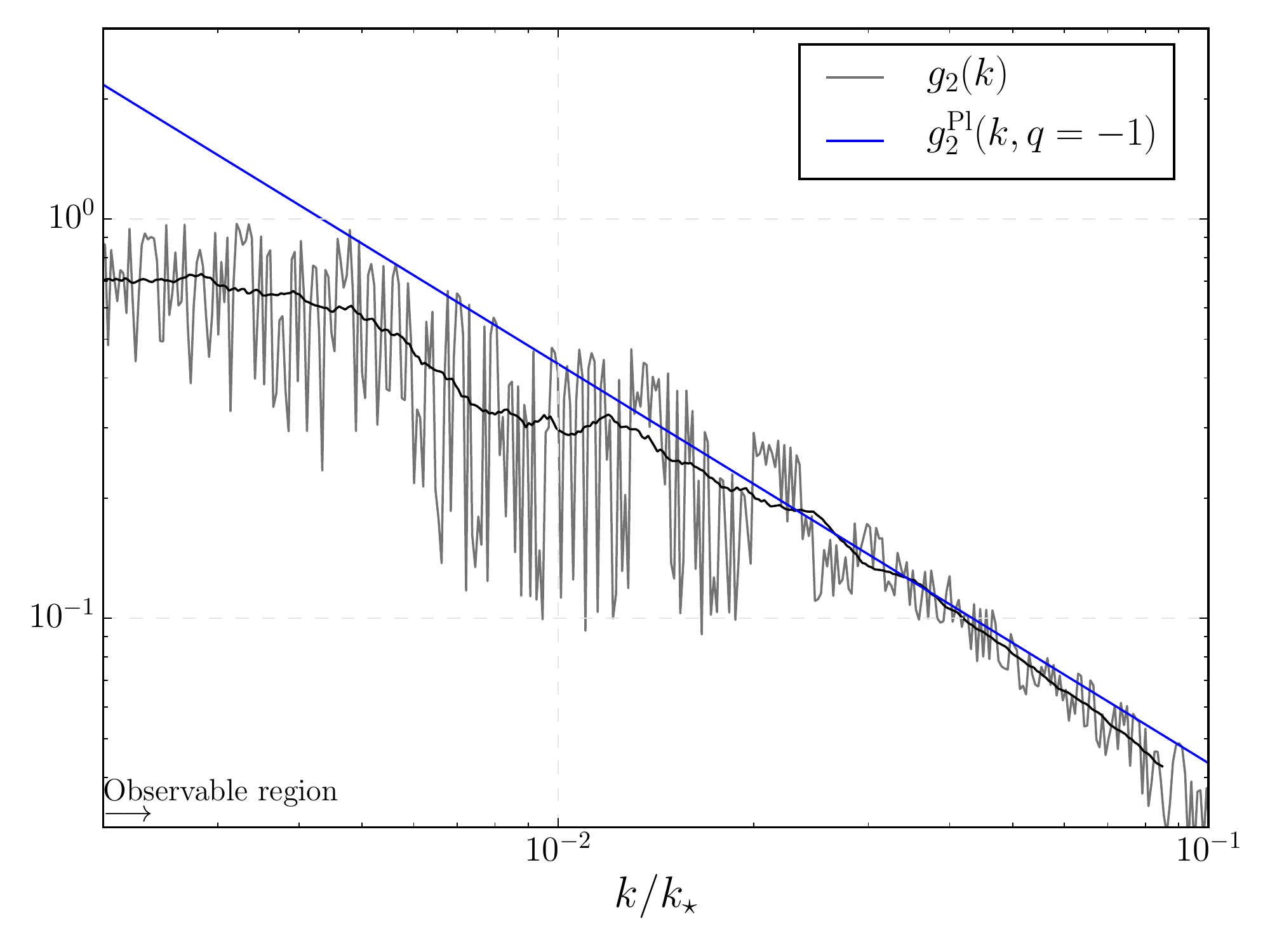}  
}
\caption{Plot of the relative amplitude $g_2(k)$ between the monopole and quadrupole (averaged over $M$) in our model (gray and black solid lines), and Planck's results  $g^{\rm Pl}_2(k)$ for $q=-1$ (blue line). The gray solid line shows our numerical results for a set of individual values of $k$, while the black solid line shows the average of the gray line, by binning it in a sufficiently small window. The outcome oscillates around the mean value with a high frequency, that is not resolved  in this plot. These oscillations do not show up in the angular correlation functions computed in the next section, since they get averaged out when integrating in
$k$. Our simulation is obtained for $\phi_{B}=1{.}1$, $\sigma^2(t_B)=5.78$ and $\sigma_{1}=0$ in Planck units.  The figure shows that the result for  $g_2(k)$  saturates Planck's constraint (the deviation is within error bars).}
\label{fig:g22M}
\end{figure}

We have performed numerical simulations  for several values of $\phi(t_B)$ and $\sigma^2(t_B)$. Here we show $\phi(t_B)=1.1$, $\sigma^2(t_B)=5.78$ and  $\sigma_{1}=0$, all in Planck units, and summarize the results in Fig.\ \ref{fig:g22M}, where we compare the amplitude of $g_2(k)$ derived from our model   with Planck's observations. We see in this figure that $g_2(k)$ falls off approximately as $1/k$  in our model, and hence we compare with Planck's results for $q=-1$. We observe that in this simulation  $g_2(k)$ saturates Planck's constraints. Either reducing  $\sigma^2(t_B)$ or increasing $\phi(t_B)$ would reduce the anisotropic features in the primordial power spectra for  scales within the observable window, and therefore the amplitude of $g_2(k)$ would also decrease. On the other hand, increasing $\sigma^2(t_B)$ up to close to the maximum value does not seem to change the results significantly, provided $\phi(t_B)$ is increased accordingly.\footnote{Our numerical analysis support this statement, although we have not been able to explore in detail what happens when $\sigma^2(t_B)$ is extremely close to the upper bound of the theory, $\sigma^2(t_B)=11.57$, since this calculation would require a prohibitively large amount of numerical resources.} Hence, the simulation considered here produces a quadrupole in the CMB compatible with Planck's observations, and picks up an approximated scale dependence of $1/k$, which fortunately is one of the parameterizations considered in \cite{plnck2018}. This simulation is therefore a good representative of the most anisotropic CMB that our model can predict without violating Planck's constraints. It is interesting to see that the simulation that reproduces the observed quadrupole has a number of $e$-folds $N$ between the bounce and the end of inflation in agreement  with the results found in \cite{barrau} for  the preferred value of $N$ in anisotropic LQC.

\subsection{Some details about  the numerics}\label{sec:numerics}

In order to deal with the dynamics of perturbations and the evaluation of the angular correlation functions, we have relied on several numerical tools. On the one hand, for the evolution of perturbations, which is the most demanding task, we have adopted the numerical tools provided by GNU scientific library. Concretely, we have used three integrators for ordinary differential equations in this library: (i) explicit embedded Runge-Kutta-Fehlberg $(4,5)$ method, (ii) explicit embedded Runge-Kutta Prince-Dormand $(8,9)$ method, and (iii) a variable-coefficient linear multistep Adams method in Nordsieck form. We always set the relative error of these algorithms equal to zero, but we vary the absolute error between $[10^{-15},10^{-17}]$. These methods give results in good agreement for the evolution of the absolute value of the basis functions ${\boldsymbol v}^{(\lambda)}_s(\vec k,t)$. However, for modes with  $(k/k_\star)\gtrsim 5\times 10^{-2}$ we observe some accumulated error in the phases of the basis functions.  This numerical error affects  the value of the cross-correlation functions  ${\cal P}_{ss'}(\v k)$ with $s\neq s'$, although only for large values of $k$ where the effects of the anisotropies are smaller. However, we have  checked that, although these numerical issues affect the very fine details of the angular correlation functions T-B and E-B discussed in the next section, the qualitative properties of these power spectra remain unaltered. Hence, the conclusions of our work for the cross-correlation functions should be understood with  this level of accuracy.

Another difficulty of these calculations has its origin in the fact that the expansion of the power spectra ${\cal P}_{ss'}(\v k)$ in spin-weighted spherical harmonics ${}_{(s-s')}Y_{L M}(\hat{k})$ requires an integral of a highly oscillatory function along the direction of $\vec k$,  $d\Omega_{\h k}=\sin\beta_kd\beta_k d\gamma_k$.  We have found that, a good balance between precision in this  integral and reasonable computational times is achieved by choosing  a uniform grid of $81\times 81$ points in the variables $u_k=\cos\beta_k\in[-1,1)$ and $\gamma_k\in[0,2\pi)$ and a 2D Simpson integration rule for the angular integral of each mode. However, this restricts ourselves to multipoles $L\leq 7$. Higher multipoles will be estimated with errors larger than 20\%. In addition, we consider around 650 points for $(k/k_\star)$ inside the interval $[10^{-3},10^{-1}]$, in a logarithmic step. For this purpose, we choose the integrator (ii) above. We suitably divided these simulations between 96 cores running simultaneously 64 Fourier modes $k$ each, and covering all possible directions in the $u_k-\gamma_k$ grid. Each  simulation, depending on the choice of absolute error,  lasts between a few days to one week. 

On the other hand, the  evaluation of the angular correlation functions requires  knowledge of the transfer functions $_s\Delta^X_{\ell}(k)$ defined in the next section. We extract these functions from CLASS (see Refs. \cite{class}). Then, we carry out an integral in the norm $k$ of the wave number. This calculation does not require large numerical resources. Therefore, we linearly interpolate the numerical estimations of ${\cal P}_{ss'}^{LM}(k)$ and $_s\Delta^X_{\ell}(k)$, and adopt a simple rectangle rule with a sufficiently high number of points, such that the final result converges. We have also  checked   that our estimations agree with those obtained by using the integration methods included in CLASS (for isotropic power spectra). 

Finally, we  observe that  increasing  $\sigma^2(t_B)$ up to a value closer to its maximum value (accompanied by a suitable increase of $\phi(t_B)$ so the anisotropic features in the power spectrum fall inside the observable window) does not introduce new physical features in the CMB, for the modes we have been able to probe. However, a larger shear increases the computational cost required to evolve modes from the initial time to the end of inflation. In addition, if spacetime anisotropies are larger, the set of modes in the observationally interesting window are more ultraviolet at the time of the bounce. This fact increases significantly the computational cost and it makes it more  challenging to maintain  numerical accuracy in our simulations. The choice $\sigma^2(t_B)=5.78$ provides a good balance between accuracy and computational time.

\section{Angular power spectrum}\label{angularps}

We have discussed in the previous section the correlation functions ${\cal P}_{ss'}(\vec{k})$ evaluated at the end of inflation. In this section we compute the impact of these primordial spectra on the CMB. More precisely, we compute the angular correlation functions of temperature anisotropies $T(\hat n)$, and the electric and magnetic components of the polarization fields, $E(\hat n)$ and $B(\hat n)$, of the CMB. 

$T(\hat n)$ and $E(\hat n)$  are real scalar fields in the  CMB sphere, while  $B(\hat n)$ is a real pseudoscalar (odd under parity). Hence, we can decompose them in angular multipoles using (zero spin-weight) spherical harmonics
\be a^X_{\ell m}=\int d\Omega\,  X(\hat n)\, \bar Y_{\ell m}(\hat n)\, ,\ \  X=T,E,B ,\ee
where the reality conditions of the fields imply $\bar a^X_{\ell m}=(-1)^m\,  a^X_{\ell -m}$. Besides, under parity, we have
\be\label{eq:alm_parity}
a_{\ell m}^{T,E}\to (-1)^\ell a_{\ell m}^{T,E},\quad \quad a_{\ell m}^{B}\to(-1)^{\ell+1} a_{\ell m}^{B}.
\ee
In this section we are interested in the correlation functions $C_{\ell \ell^{\prime},m m^{\prime}}^{X, X^{\prime}}=\left\langle a_{\ell m}^{X} a_{\ell^{\prime} m^{\prime}}^{X^{\prime} }\right\rangle$. The fields $T(\hat n)$,  $E(\hat n)$, and $B(\hat n)$ are sourced by the primordial perturbations $\Gamma_s$ that we have discussed in the previous sections. The relation between them is found by evolving the fields  $\Gamma_s$ across the radiation dominated era, and then computing their effects on the temperature and polarization anisotropies of the CMB. The complex physics involved in this process is encoded in the so-called transfer functions $_s\Delta^X_{\ell}(k)$. More concretely,  these functions relate the value of $\Gamma_s(\vec k)$ at the end of inflation with the angular multipoles $a^X_{\ell m}$ by means of  
\be \label{alm} a^X_{\ell m}=\int \frac{d^3 k}{(2\pi)^3}\, (-i)^{\ell}\, \sum_{s=0,\pm2} {_s\Delta}^X_{\ell}(k)\, \Gamma_s(\vec k)\,  {_s\bar{Y}}_{\ell m}(\hat k)\, . \ee
The functions  $_s\Delta^X_{\ell}(k)$ can be computed, for instance, using a Boltzmann code such as CLASS (see Refs. \cite{class}). The well known fact that scalar perturbations $\Gamma_0$ do not generate B-polarization in the CMB, is reflected in the fact that $_0\Delta^B_{\ell}(k)=0$. Furthermore, the transformation properties under parity of $T(\hat n)$, $E(\hat n)$ and $B(\hat n)$ imply
\be _{-s}\Delta^T_{\ell}(k)=\, _s\Delta^T_{\ell}(k) \, , \ \ \ \ _{-s}\Delta^E_{\ell}(k)=\, _s\Delta^E_{\ell}(k) \, , \ \ \ \ _{-s}\Delta^B_{\ell}(k)=\, -\, _s\Delta^B_{\ell}(k)\, . \ee
Moreover, they remain invariant under  inversions $\v k\to - \v k$, since $_s\Delta^X_{\ell}(k)$ only depend on the norm of $\vec k$ and not on its direction (recall that the universe after inflation is extremely isotropic). We will use these properties in the rest of this section. 

Expressions (\ref{alm}) can be used to write the correlation functions $C_{\ell \ell^{\prime},m m^{\prime}}^{X, X^{\prime}}=\left\langle a_{\ell m}^{X} a_{\ell^{\prime} m^{\prime}}^{X^{\prime} }\right\rangle$ in terms of the primordial power spectra ${{\cal P}}_{ss'}(\vec{k})$ , $s,s'=0, \pm 2$, as
\be C_{\ell \ell^{\prime}, m m^{\prime}}^{X, X^{\prime}}=\int \frac{d^3 k}{(2\pi)^3}\, (-i)^{(\ell+\ell')} \, \sum_{s,s'} \, {_s\Delta}^X_{\ell}(k)\, {_{s'}\Delta}^{X'}_{\ell'}(k)\,\frac{2\pi^2}{k^3}{{\cal P}\,}_{ss'}(\vec{k})\,  {_s\bar{Y}}_{\ell m}(\hat k)\,  {_{s'}\bar{Y}}_{\ell' m'}(-\hat k)\, . \ee
These expressions are all that we need to compute the predictions for the CMB from the results of the previous section. Notice that the invariance under parity of the primordial spectra ${{\cal P}\,}_{ss'}(\vec{k})$, implies that the angular correlation functions $ C_{\ell \ell^{\prime}, m m^{\prime}}^{X, X^{\prime}}$ are also parity-invariant. Using  (\ref{eq:alm_parity}), this implies 
\bea  C_{\ell \ell^{\prime}, m m^{\prime}}^{TT}&=&C_{\ell \ell^{\prime}, m m^{\prime}}^{EE}=C_{\ell \ell^{\prime}, m m^{\prime}}^{BB}=C_{\ell \ell^{\prime}, m m^{\prime}}^{TE}=0\, \  \ \  \ \ {\rm if} \ \ell+\ell^{\prime}\, {\rm odd}\, , \\ 
 C_{\ell \ell^{\prime}, m m^{\prime}}^{TB}&=& C_{\ell \ell^{\prime}, m m^{\prime}}^{EB}=0\, \  \ \  \ \ {\rm if} \ \ell+\ell^{\prime}\, {\rm even} \, .\ea
In FLRW spacetimes, isotropy further implies that all angular correlation functions vanish unless $\ell=\ell'$. Therefore, parity combined with isotropy implies  $C_{\ell \ell^{\prime}, m m^{\prime}}^{TB}= C_{\ell \ell^{\prime}, m m^{\prime}}^{EB}=0$ for all $\ell$ and $\ell'$. But in Bianchi I spacetimes, these cross-correlations can be different from zero for $\ell+\ell'$ equal to an odd number. Hence, the presence of these correlations in the CMB is a smoking gun  for anisotropies. The value of these cross-correlations and the concrete way they vary with   $\ell$ and $\ell'$ is one of the most important predictions of our model.

We show now the results for all angular correlation functions for the Bianchi I solution  discussed in the previous section, that corresponds to $\phi_{B}=1{.}1$, $\sigma^2(t_B)=5.78$ and $\sigma_{1}(t_B)=0$, all in Planck units. (Recall that in this solution there are $N=70.1$ $e$-folds of expansion between the bounce and the end of inflation.)

\begin{enumerate}

\item {\bf T-T angular correlation function.}
As we just mentioned, these correlations are different from zero only for even $\ell+\ell'$. All primordial power spectra ${{\cal P}}_{ss'}(\vec k)$, with $s,s'=0,\pm2$, contribute to  $C_{\ell \ell^{\prime}, m m^{\prime}}^{TT}$, although only with  even multipoles $L$. We plot in the left panel of Figure \ref{fig:DTT} the  angular correlation function for temperature-temperature anisotropies $D_{\ell}^{TT}$, defined as 
\be D_{\ell}^{TT}\equiv\frac{T_0^2 \ell (\ell+1)}{2\pi}\,  C_{\ell}^{TT}\, , \ee
where $C_{\ell}^{TT}=\frac{1}{2\ell+1} \sum_{m=-\ell}^\ell (-1)^mC_{\ell \ell, m -m}^{TT}$. We also include, for comparison, $\mathring D_{\ell}^{TT}$, the angular correlation function obtained from an almost scale-invariant isotropic primordial spectrum, together with the uncertainty in observations due to cosmic variance. We observe that our model predicts a modest enhancement  of correlations  at low multipoles, although too small to be distinguished from the standard predictions once  cosmic variance is taken into account. In the right panel, we plot the anisotropic angular correlation function
\be D_{\ell \ell+2, 00}^{TT}\equiv \frac{T_0^2 \ell (\ell+1)}{2\pi}\,  C_{\ell \ell+2, 00}^{TT}\, , \ee
together with the result from the isotropic case (which vanishes identically). For this particular choice of Bianchi-I geometry, these off-diagonal components of the T-T angular  correlation function are negative. Its magnitude is large for low multipoles $\ell$, and then decreases as $\ell$ increases, as expected. This result  is compatible with Planck's observations of a quadrupolar modulation in the temperature map \cite{plnck2018}.
\begin{figure}[h]
{\centering     
\includegraphics[width = 0.49\textwidth]{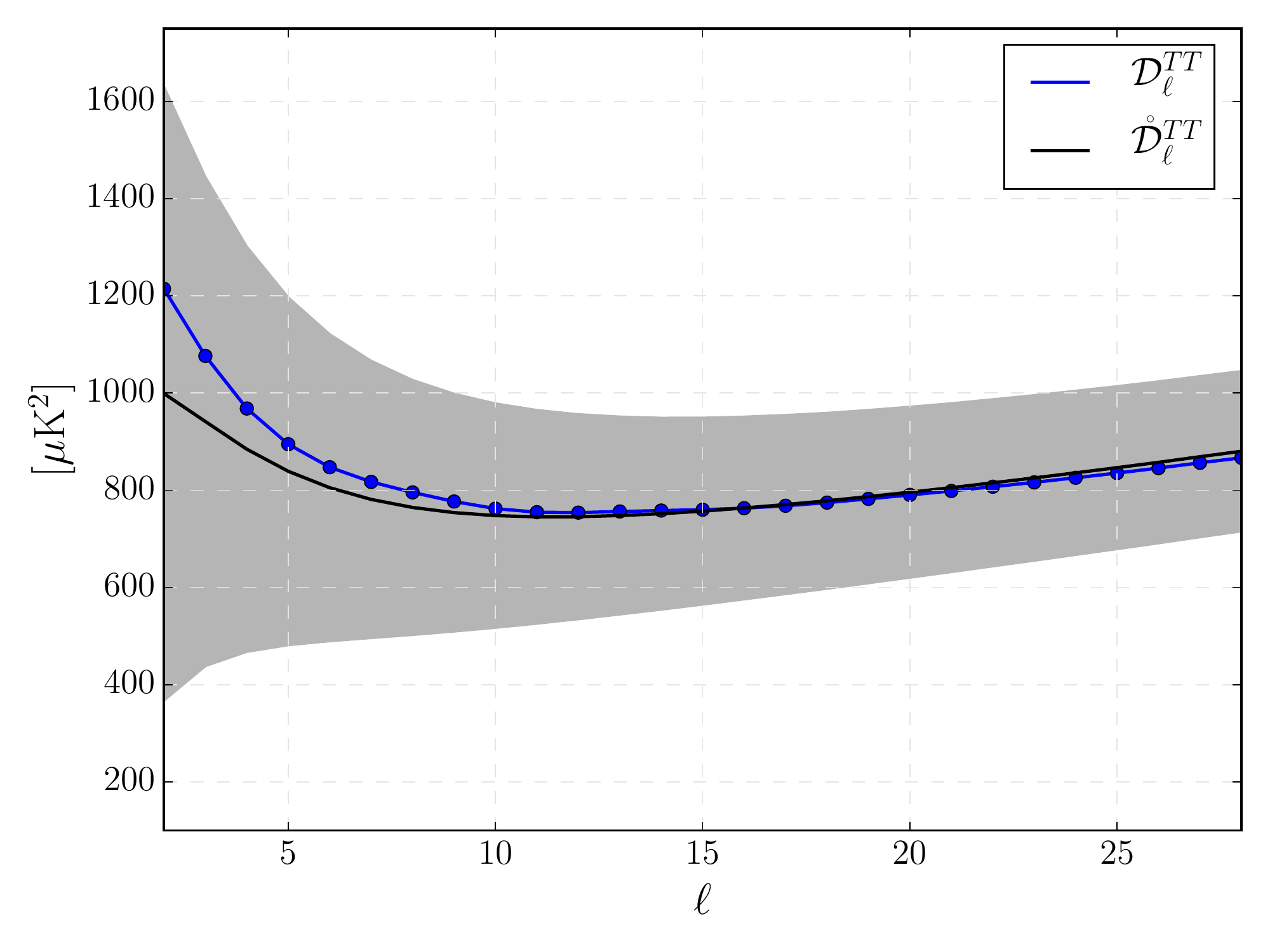}
\includegraphics[width = 0.49\textwidth]{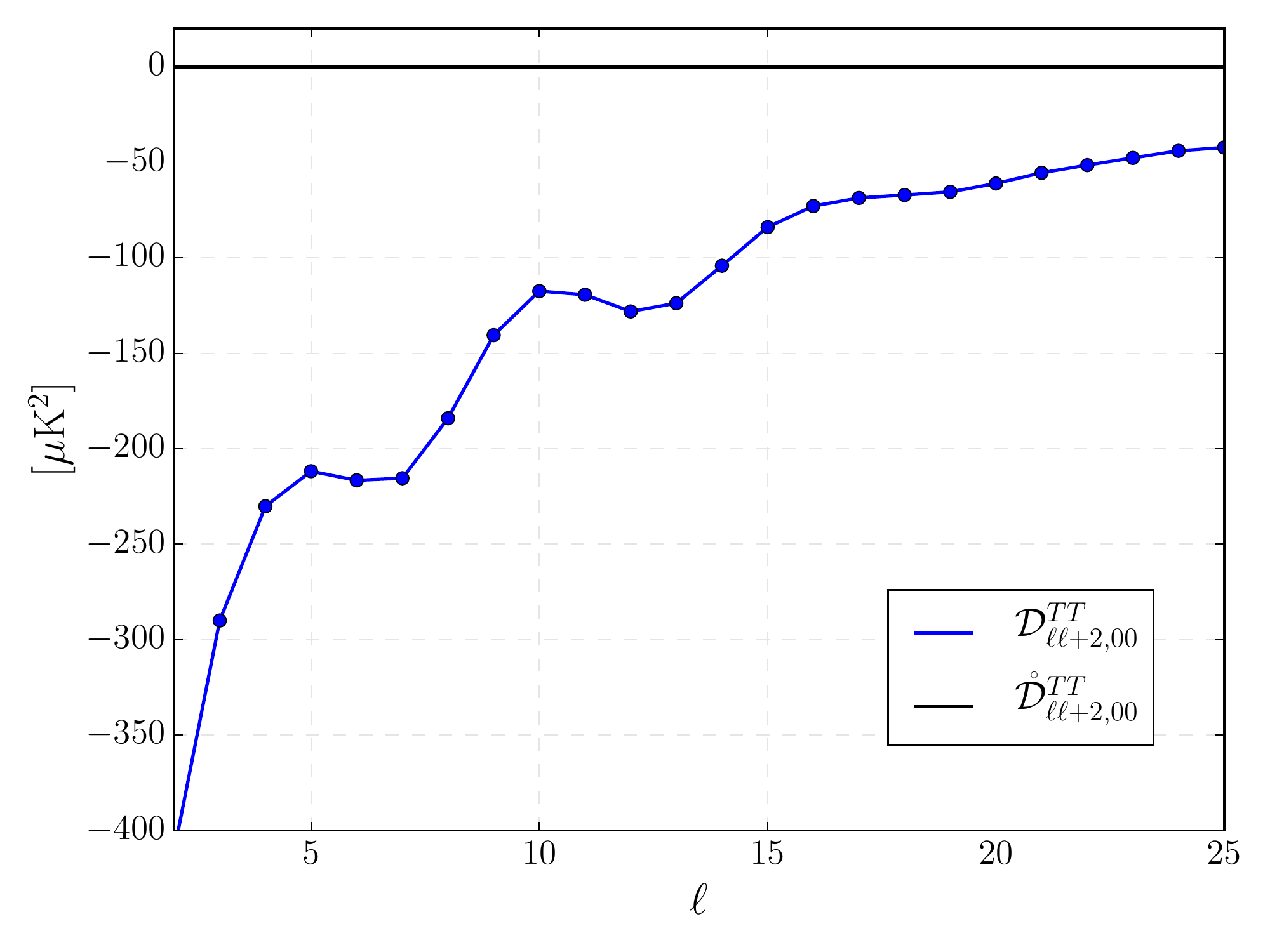}  
}
\caption{Left panel: Plot of the temperature-temperature angular correlation function $D_{\ell}^{TT}$. We also show $\mathring D_{\ell}^{TT}$, obtained  from an almost  scale-invariant isotropic primordial spectrum, for comparison. The difference of these two correlation functions is smaller than  the uncertainties coming from cosmic variance.
Right panel: We plot the temperature-temperature angular correlation function $D_{\ell \ell+2, 00}^{TT}$ along with its isotropic counterpart, which is zero.}
\label{fig:DTT}
\end{figure}

\item {\bf E-E  correlation function.}
These correlations are similar to the previous ones, in the sense that they  are different from zero only for even $\ell+\ell'$, and in that  all primordial spectra ${{\cal P}}_{ss'}(\vec k)$ with $s,s'=0,\pm2$ contribute, although only with even multipoles. We plot in Fig. \ref{fig:DEE}, the  $\ell=\ell'$ component versus $\ell$. More precisely, we plot
\be D_{\ell}^{EE}\equiv \frac{T_0^2 \ell (\ell+1)}{2\pi}\,  C_{\ell}^{EE}\, , \ee
with $C_{\ell}^{EE}=\frac{1}{2\ell+1} \sum_{m=-\ell}^\ell (-1)^mC_{\ell \ell, m -m}^{EE}$. We also plot the results for a scale-invariant isotropic primordial spectrum $\mathring D_{\ell}^{EE}$ for comparison. The conclusion are the same as for the temperature-temperature correlations. Namely, there is a small  enhancement of power at low multipoles. Besides, off-diagonal components of the angular power spectrum are different from zero, with considerable more power at low multipoles. As an example, we show in the right panel of Fig. \ref{fig:DEE} the angular correlation function
\be D_{\ell \ell+2, 00}^{EE}=\frac{T_0^2 \ell (\ell+1)}{2\pi}\,  C_{\ell \ell+2, 00}^{EE}\, . \ee
For the particular  Bianchi-I geometry chosen in this simulation, these off-diagonal components of the E-E angular correlation function are negative. Their magnitude decrease as $\ell$ increases.  
\begin{figure}[h]
{\centering     
\includegraphics[width = 0.49\textwidth]{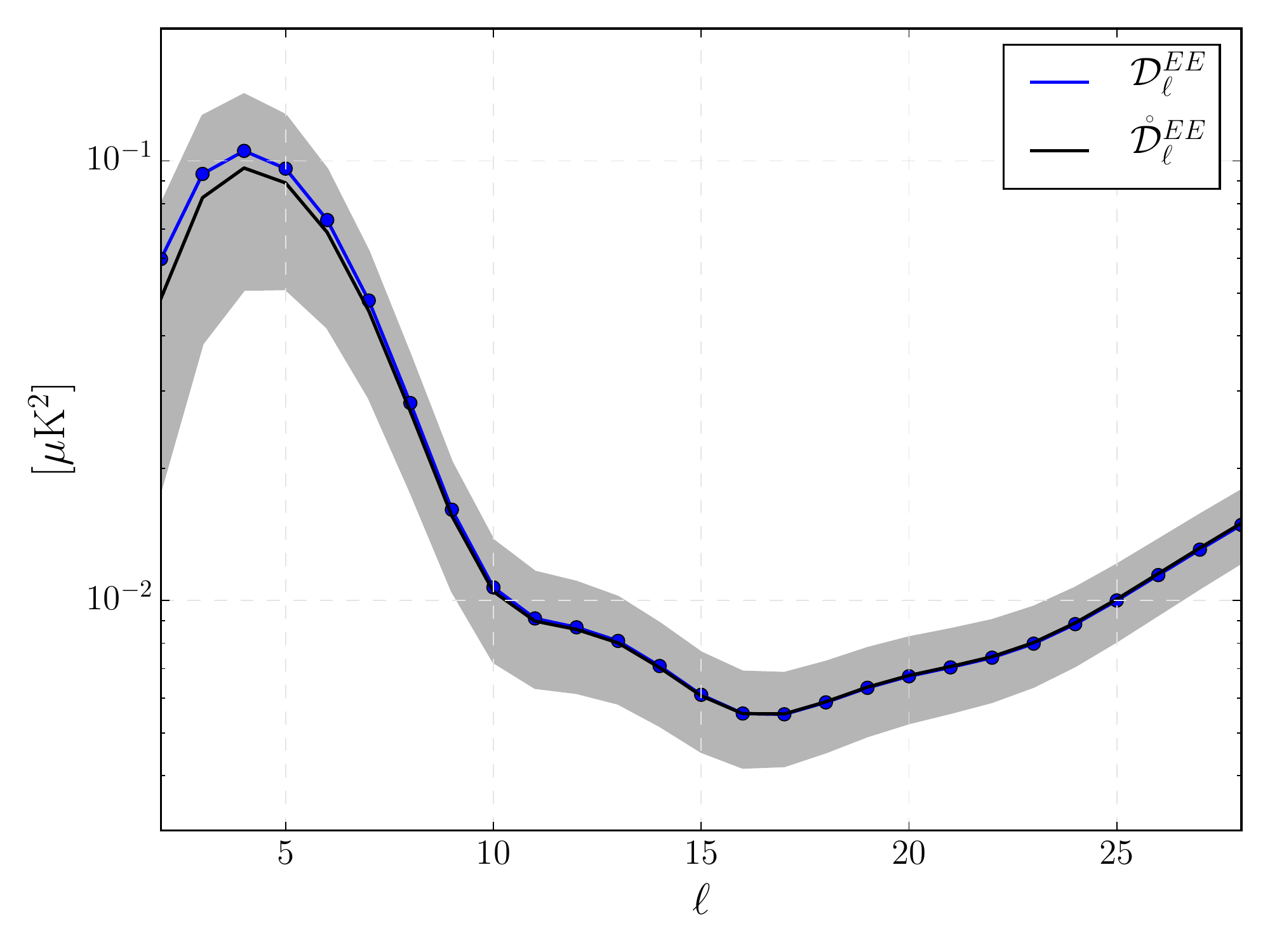}  
\includegraphics[width = 0.49\textwidth]{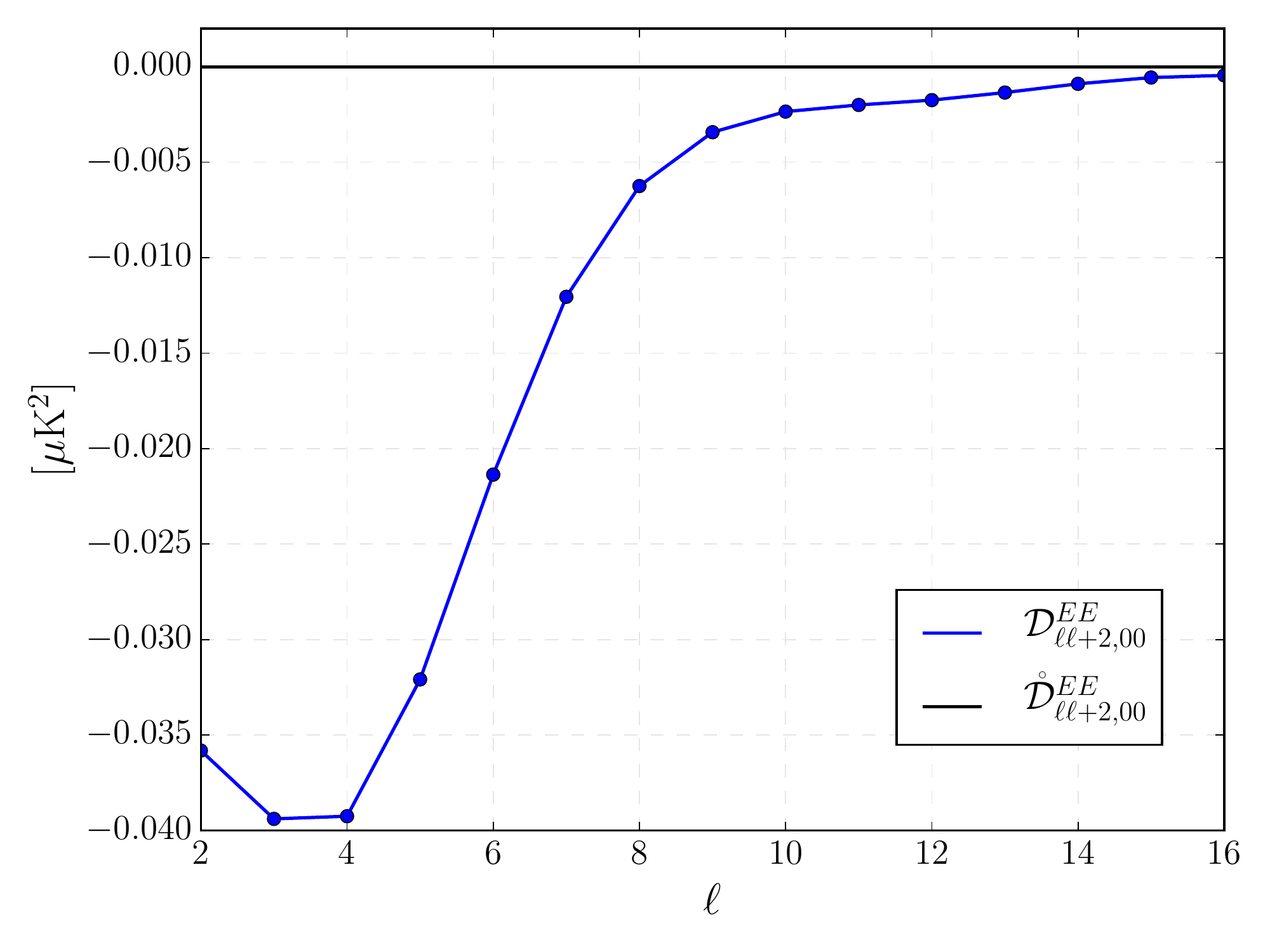} 
}
\caption{Left panel: Comparison of our E-E polarization angular correlation function $D_{\ell}^{EE}$ and the one obtained  from a scale-invariant isotropic primordial spectrum, denoted by $\mathring D_{\ell}^{EE}$. Their difference is smaller than  the uncertainties coming from cosmic variance. Right panel: We show another non vanishing component of the E-E polarization correlation function $D_{\ell \ell+2, 00}^{EE}$, and its counterpart obtained from an isotropic primordial power spectrum, which is exactly zero. }
\label{fig:DEE}
\end{figure}

\item {\bf T-E cross-correlation function.}
  The T-E angular cross-correlations share the properties of the two previous cases. Let us define again the average $C_{\ell}^{TE}=\frac{1}{2\ell+1} \sum_{m=-\ell}^\ell (-1)^mC_{\ell \ell, m -m}^{TE}$. In the left panel of Fig. \ref{fig:DTE} we show
\be D_{\ell}^{TE}\equiv \frac{T_0^2 \ell (\ell+1)}{2\pi}\,  C_{\ell}^{TE}\, , \ee
together with its counterpart obtained  from a nearly scale-invariant isotropic primordial spectrum $\mathring D_{\ell}^{TE}$. The conclusions are the same as for the T-T and E-E correlations. There is an enhancement of power at low multipoles, but not significantly enough once cosmic variance is taken into account.

Additionally, the right panel of Fig. \ref{fig:DEE} contains a plot of the off-diagonal angular correlation function
\be D_{\ell \ell+2, 00}^{TE}\equiv \frac{T_0^2 \ell (\ell+1)}{2\pi}\,  C_{\ell \ell+2, 00}^{TE}\, . \ee
This correlation function is zero in the isotropic case. We see in Fig. \ref{fig:DTE} that the  amplitude of $D_{\ell \ell+2, 00}^{TE}$ decreases for large $\ell$.

\begin{figure}[h]
{\centering     
\includegraphics[width = 0.49\textwidth]{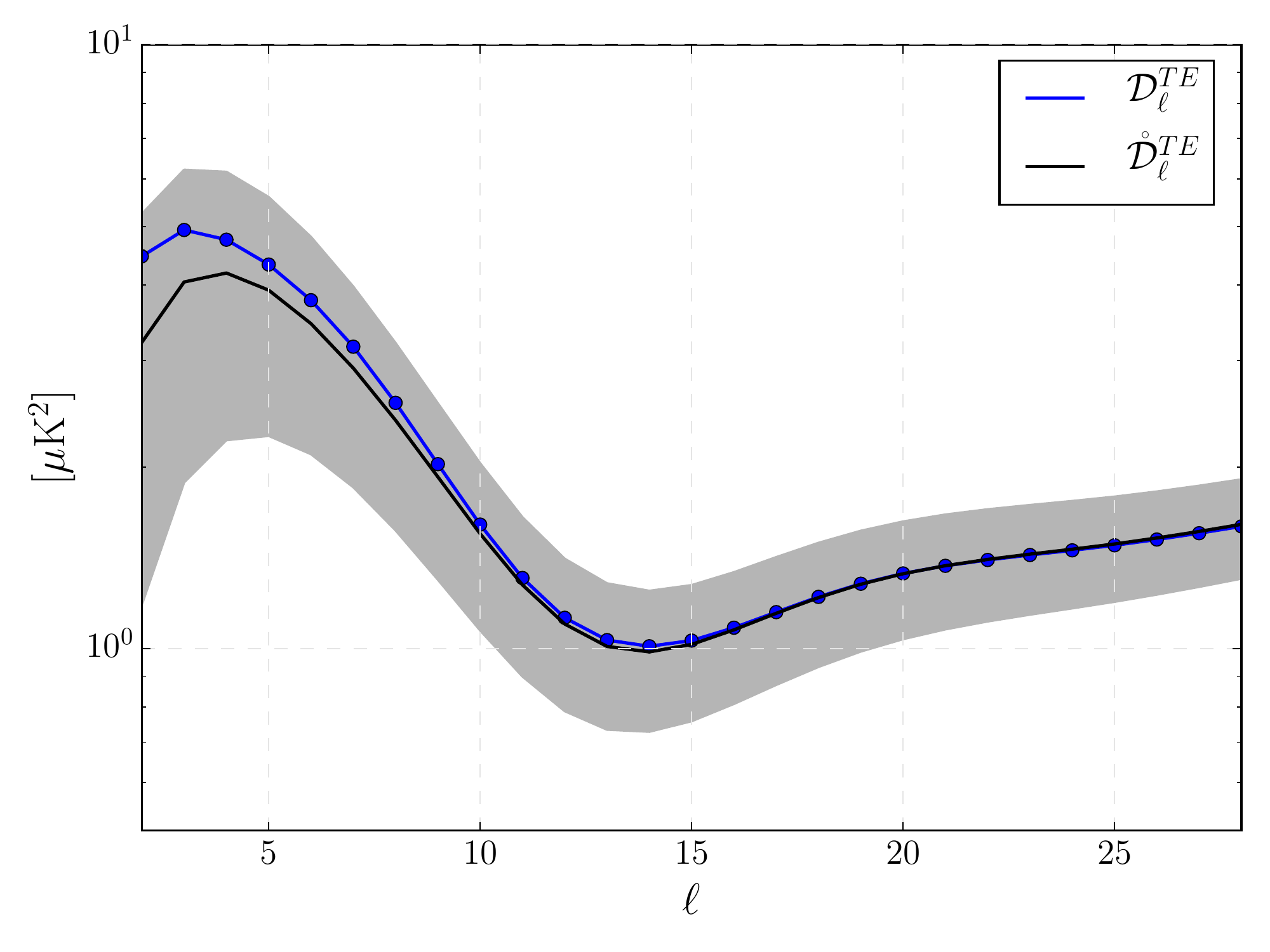}  
\includegraphics[width = 0.49\textwidth]{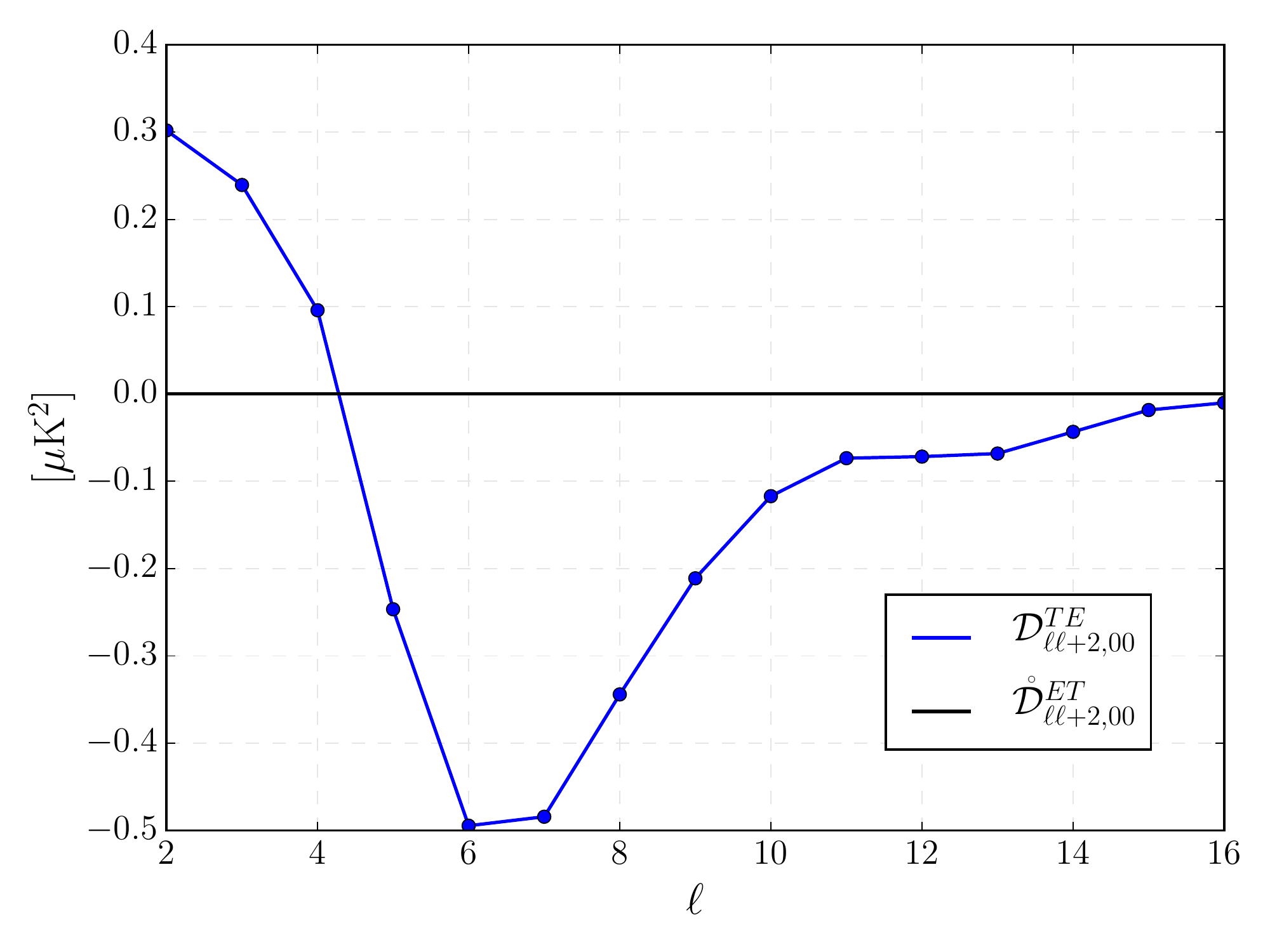} 
}
\caption{Left panel: T-E cross-correlation  $D_{\ell}^{TE}$, and  its counterpart  obtained  from a scale-invariant isotropic primordial spectrum, $\mathring D_{\ell}^{TE}$. The difference between them is smaller than  the uncertainties coming from cosmic variance. Right panel: We show one of the nonvanishing components of the T-E polarization correlation function, concretely  $D_{\ell \ell+2, 00}^{TE}$, and its isotropic counterpart $\mathring D_{\ell \ell+2, 00}^{TE}=0$.}
\label{fig:DTE}
\end{figure}

\item {\bf B-B  correlation function.}
Again, for this angular correlation function $\ell+\ell'$ must be even, otherwise $C_{\ell \ell', m m^{\prime}}^{BB}$ vanishes. Furthermore, as mentioned above, only the purely tensorial primordial spectra ${{\cal P}}_{\pm 2\pm 2}(\vec k)$ and ${{\cal P}}_{\mp 2\pm 2}(\vec k)$ contribute, and  only with even multipoles $L$. Their lowest multipoles are $L=0$ for ${{\cal P}}_{\pm 2\pm 2}(\vec k)$, and $L=4$ for ${{\cal P}}_{\mp 2\pm 2}(\vec k)$. 

We plot $D_{\ell}^{BB}$ in the left panel of Fig. \ref{fig:DBB}, defined as
\be D_{\ell}^{BB}\equiv \frac{T_0^2 \ell (\ell+1)}{2\pi}\,  C_{\ell}^{BB}\, , \ee
where $C_{\ell}^{BB}=\frac{1}{2\ell+1} \sum_{m=-\ell}^\ell (-1)^mC_{\ell \ell, m -m}^{BB}$. Once more, we also show the result obtained from a nearly scale-invariant isotropic primordial spectrum, $\mathring D_{\ell}^{BB}$, for comparison. Figure \ref{fig:DBB} shows that their differences at low multipoles are larger than in  previous cases, although  still  small compared to  cosmic variance. In the right panel of Fig. \ref{fig:DBB}, we also show some of the off-diagonal elements of this correlation function. Concretely,
\be D_{\ell \ell+2, 00}^{BB}\equiv \frac{T_0^2 \ell (\ell+1)}{2\pi}\,  C_{\ell \ell+2, 00}^{BB}\, . \ee
We see that this quantity is large only at low multipoles, and then decreases, reaching values compatible with zero for $\ell\neq\ell'$, for $\ell\gg 10$. 
\begin{figure}[h]
{\centering     
\includegraphics[width = 0.49\textwidth]{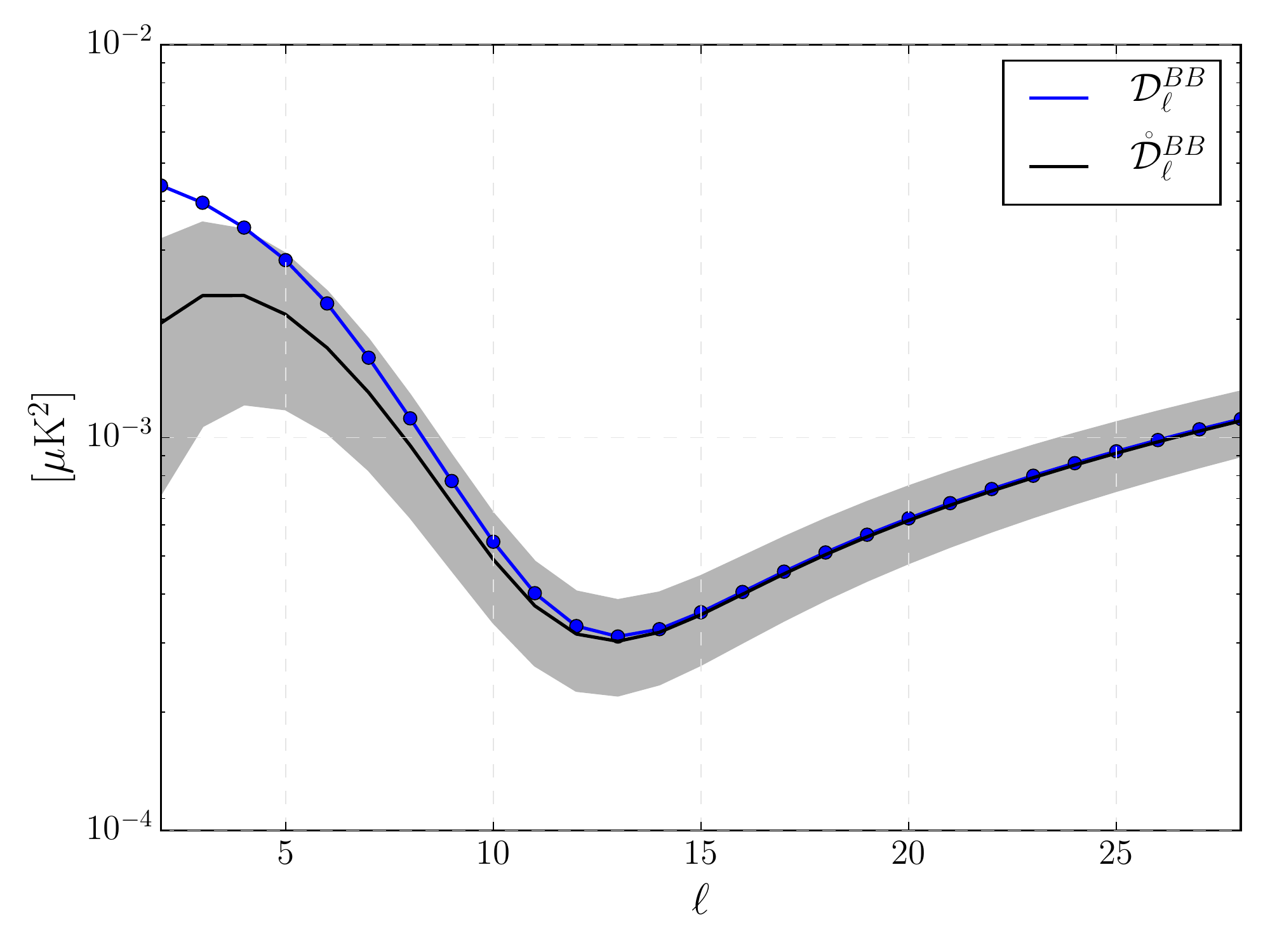}  
\includegraphics[width = 0.49\textwidth]{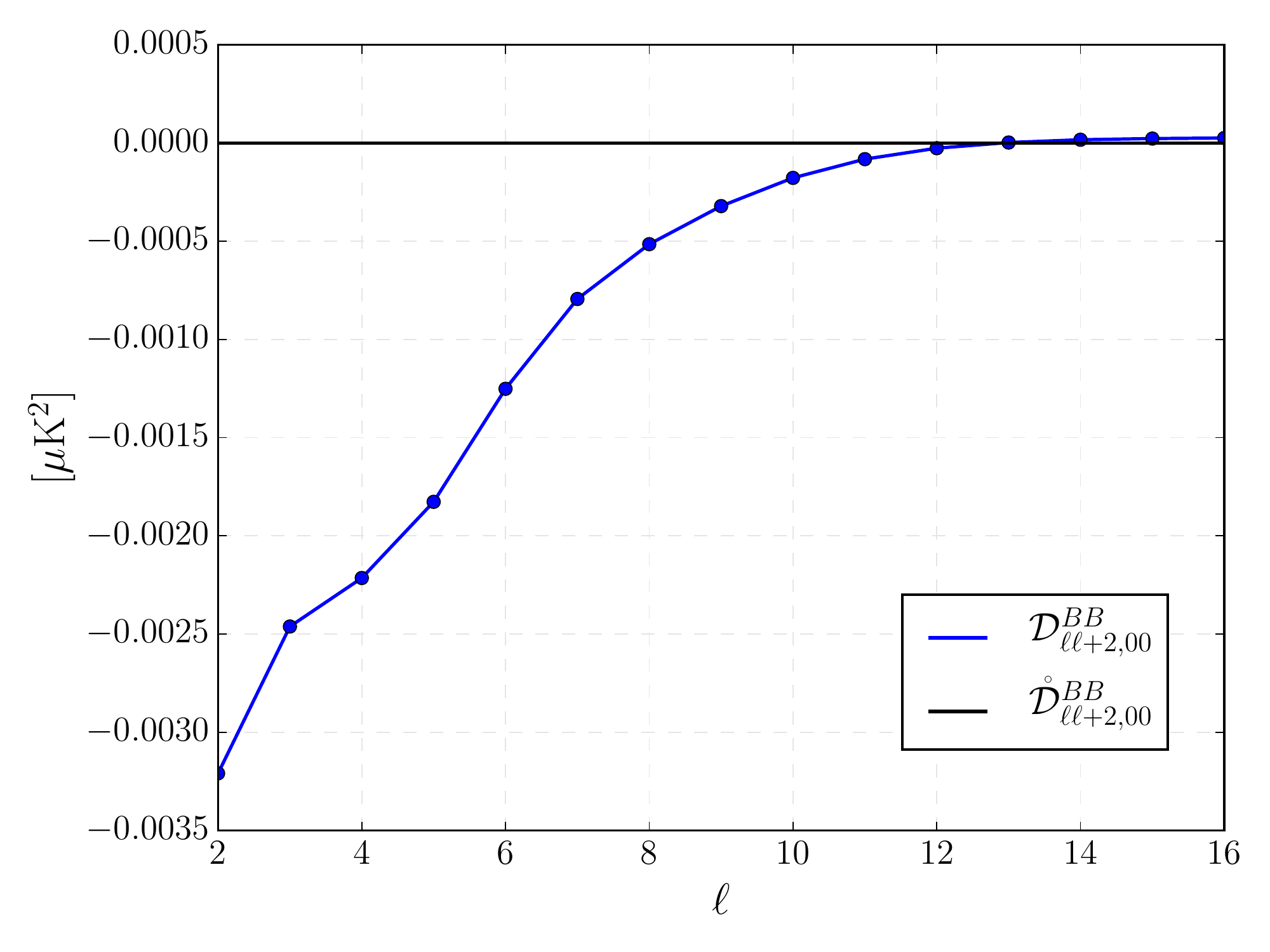} 
}
\caption{Left panel: B-B polarization angular correlation function $D_{\ell}^{BB}$. On the other hand, $\mathring D_{\ell}^{BB}$ is obtained  from a scale-invariant isotropic primordial spectrum, and it is shown for comparison. We observe a stronger enhancement at low multipoles than for T-T and E-E correlations, but not significant enough. Right panel: Off-diagonal component of the B-B polarization correlation function $D_{\ell \ell+2, 00}^{BB}$. Once more, we also show the  same correlation function computed from an isotropic primordial power spectrum, $\mathring D_{\ell \ell+2, 00}^{BB}=0$, for comparison.}
\label{fig:DBB}
\end{figure}

\item {\bf T-B and E-B   correlations.}
Only odd values of $\ell+\ell'$  produce a non-zero result. Furthermore, the result comes entirely from  the primordial cross-correlations   ${{\cal P}}_{\pm 2\mp 2}(\vec k)$, ${{\cal P}}_{0\pm 2}(\vec k)$ and ${{\cal P}}_{\pm 2 0}(\vec k)$, although only with odd multipoles $L$. We plot in Fig. \ref{fig:TB} the correlation functions 
 
\be D_{\ell}^{TB}=\frac{T_0^2 \ell (\ell+1)}{2\pi}\,  C_{\ell \ell+1, 0 0}^{TB}\, , \ee
and 
\be D_{\ell}^{EB}=\frac{T_0^2 \ell (\ell+1)}{2\pi}\,  C_{\ell \ell+1, 0 0}^{EB}\, , \ee
where we have chosen $m=m'=0$ as a representative case. These correlations are identically zero in the isotropic scenario, but they are not in our model and, as in previous cases, they reach their largest amplitudes at low multipoles. They oscillate around zero, and their amplitude decreases considerably for $\ell\gg 10$. 
\begin{figure}[h]
{\centering     
  \includegraphics[width = 0.49\textwidth]{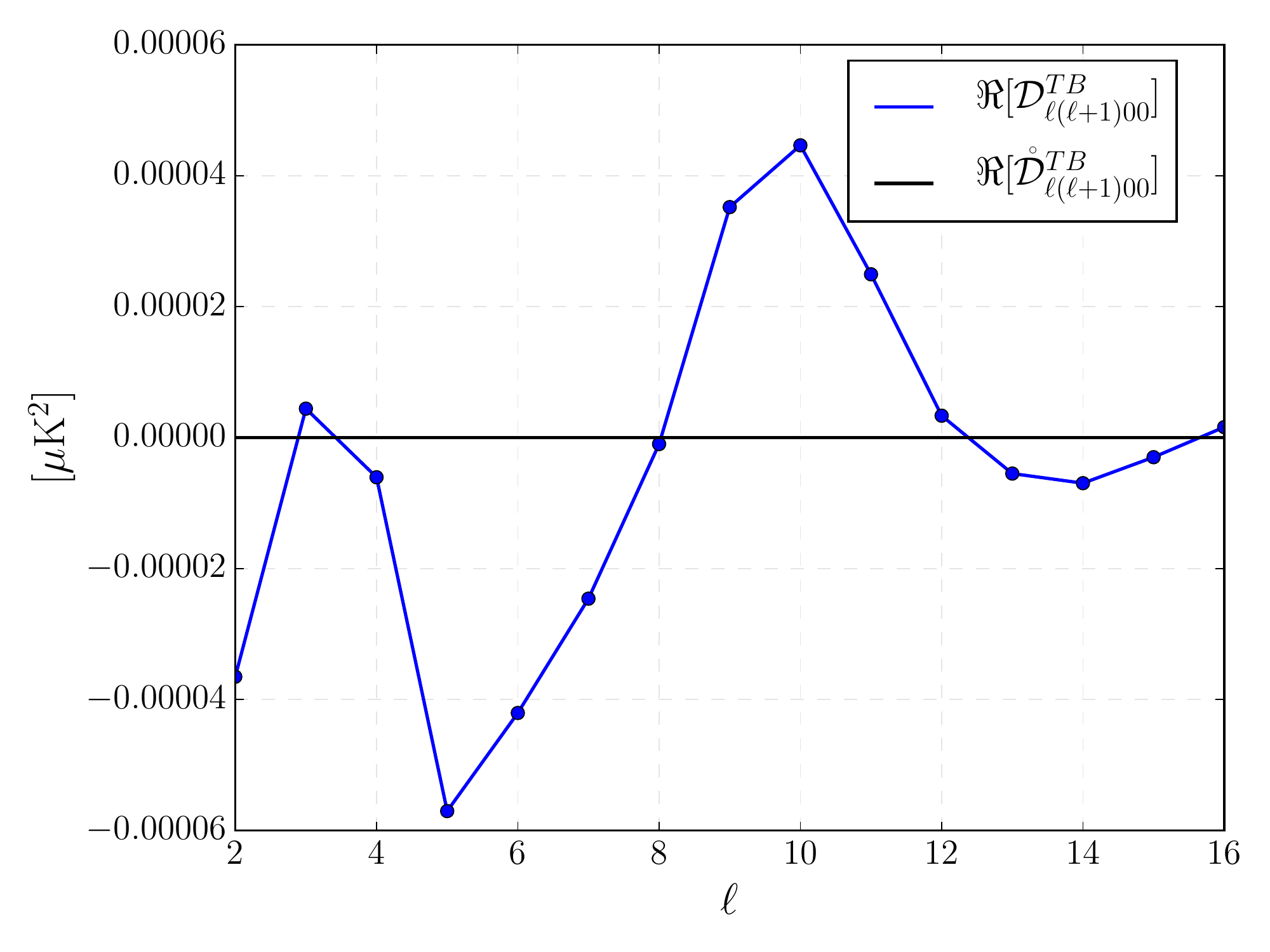}
  \includegraphics[width = 0.49\textwidth]{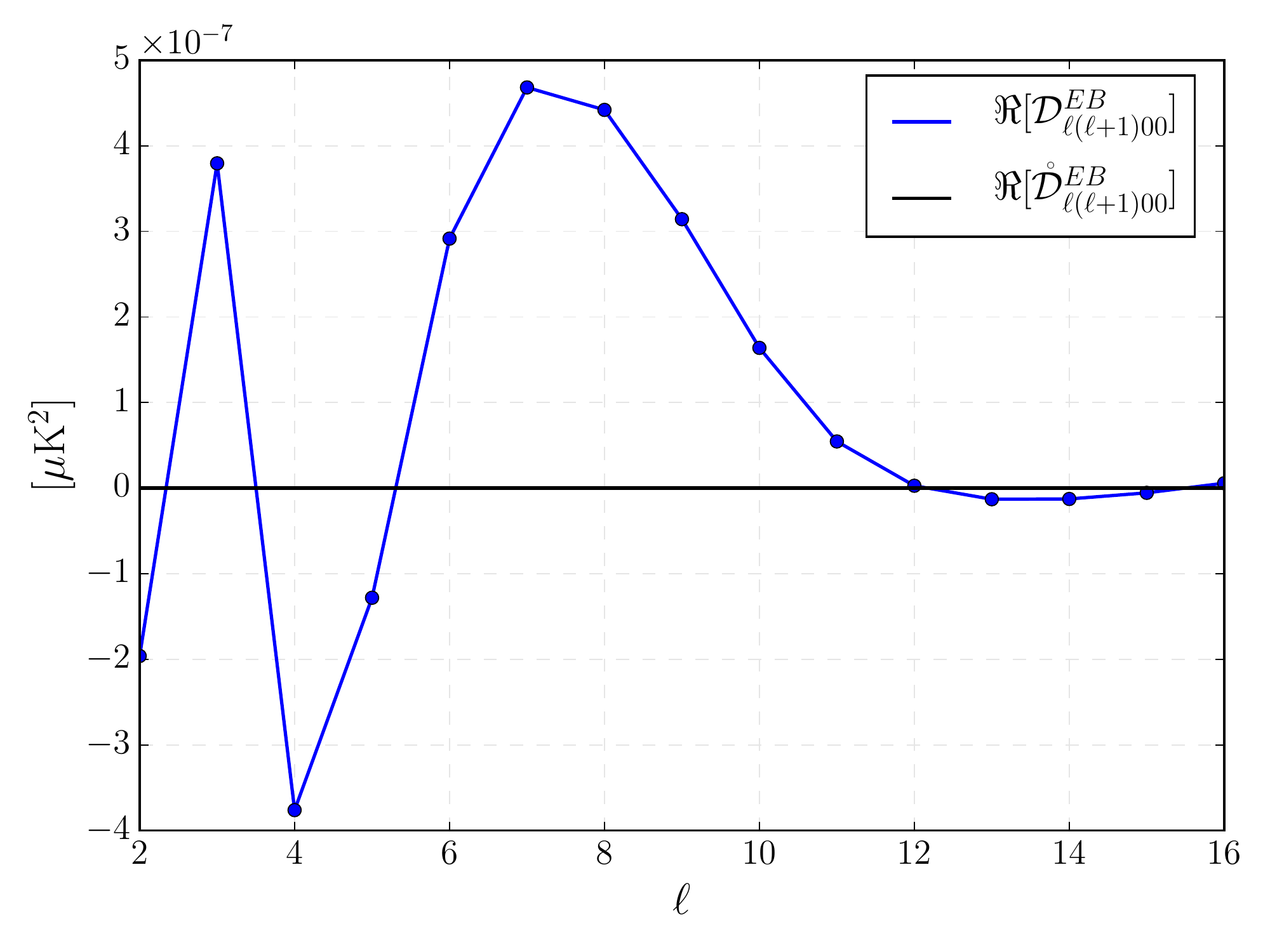}  
}
\caption{ T-B (left panel) and E-B (right panel)  angular correlation functions at low multipoles, and the isotropic counterpart for comparison.}
\label{fig:TB}
\end{figure}

\end{enumerate}

In summary, we have seen that the effects of an anisotropic bounce that is compatible with the observed quadrupole, produces a modest enhancement of power in all the diagonal correlation functions  (i.e.\ all nonvanishing $C^{XX'}_{\ell \ell', mm'}$ with $\ell=\ell'$, $m=-m'$) at  low multipoles, although  not significantly enough when compared with the effects of cosmic variance. However, off-diagonal elements of these correlation functions are different from zero at low multipoles.  We also observe a larger enhancement in B-polarization than in  temperature and E-modes. 
This feature has its origin in the coupling between scalar and tensor modes caused by anisotropies, as well as from the fact that tensor modes are more sensitive to anisotropies than the scalar perturbations. Hence, if B-modes are measured in the near future (see e.g. \cite{core}), our predictions could be tested. Some of the anisotropic correlations and cross-correlations that we predict are small, and probably  difficult to be observed. But others are not, and can be used to test our ideas. 

It is important to emphasize that the results of this section rest on a choice of potential $V(\phi)$, and also on a selection of the rest of  cosmological parameters, which are needed to compute the angular correlation functions. We end this section with a discussion about  the consequences of these choices. 

Regarding the potential $V(\phi)$ for the scalar field, the previous plots have been obtained using a quadratic potential $V(\phi)=\frac{1}{2} m^2\phi^2$, with the value $m$ that best fits CMB data  \cite{plnck2018}, namely  $m=1.28\times10^{-6}$ in Planck units. This choice fixes the spectral index of scalar perturbations $n_s$, and the amplitude of tensor modes. However, because the anisotropies at low multipoles originate from physics that is independent of $V(\phi)$, the  anisotropic features  described above do not depend on our choice; except for correlations involving B-polarization, since their overall amplitude depends on  $V(\phi)$. Hence, our invariant prediction for correlation functions involving B-modes is the  amplitude of the anisotropic features {\it relative} to the overall amplitude.

On the other hand, we have used the values of the rest of cosmological parameters ($\Omega_b$, $\Omega_c$, $\theta_{MC}$, $\tau$) that were reported in Ref. \cite{plnck2018}. But it is important to keep in mind that these values are derived in \cite{plnck2018} by means of a Bayesian analysis {\it that assumes an isotropic and almost scale invariant} primordial spectrum of scalar perturbations. Since our primordial scalar spectra is neither isotropic nor  scale invariant for small values of $k$, we should be concerned about the self-consistency of this strategy. We  have analyzed this question and concluded that our calculation is in fact self-consistent,  because the mean values of the marginalized cosmological parameters are quite insensitive to the anisotropic modifications that our model introduces in the primordial spectra. The intuitive reason for this is because, as the plots above show, our model produces a very small modification to the isotropic correlation functions $\mathcal{D}^{TT}_{\ell}$, $\mathcal{D}^{EE}_{\ell}$, and $\mathcal{D}^{TE}_{\ell}$, from which the cosmological parameters are obtained. And, on the other hand, because the new physics in our model is restricted to low multipoles $\ell \lesssim 30$,  and their statistical weight is small relative to the rest of multipoles $\ell\in [30,2500]$. Therefore, we find that there is a neat separation between the physics during inflation and later times, that determines the  best-fit values of the cosmological parameters, and the new physics that our model introduces, which affects mainly anisotropic correlations at low multipoles $\ell$. 

\begin{figure}[h]
{\centering     
\includegraphics[width = 1.0\textwidth]{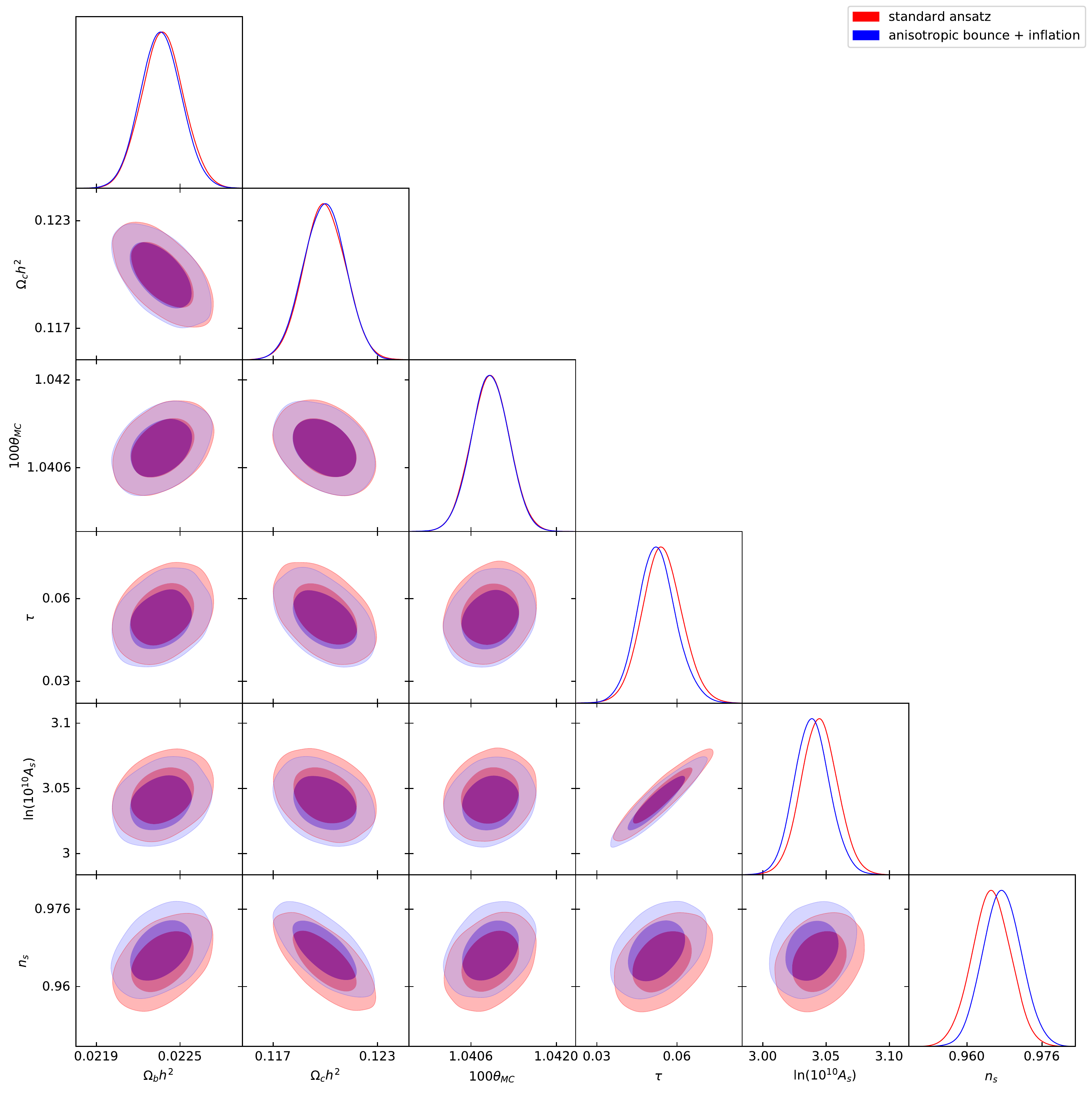}}
\caption{Comparison of the $68\%$ and $95\%$ probability contours for the six parameters $A_s$, $n_s$, $\Omega_b$, $\Omega_c$, $\theta_{MC}$, and $\tau$, of the $\Lambda$CDM model with the standard ansatz of an almost scale invariant spectrum of primordial perturbations, and LQC where an anisotropic bounce takes place before the inflationary era. 
The mean values of both distributions are well within 1 standard deviation of each other.}
\label{plot}
\end{figure}

In order to be more quantitative about these statements, we have used a Markov chain Monte Carlo analysis (MCMC), using TT, EE, TE, and lensing  data to find the best fit to the six free cosmological parameters $\Omega_b$, $\Omega_c$, $\theta_{MC}$, $\tau$, $A_s$ and $n_s$---note that we have not fixed a potential $V(\phi)$ in this analysis, but we have rather parametrized the freedom in the choice of $V(\phi)$ by means of  the amplitude of the scalar primordial spectrum  $A_s$ and its spectral index $n_s$. We have obtained that  the mean values of the six parameters are very close to the values obtained without the anisotropic bounce. Figure \ref{plot} shows the 1 and 2-sigma confidence contours for the six parameters. The mean values of marginalized posterior distributions of these parameters are well within 1-sigma of the ones obtained from isotropic inflation. The largest deviation from the corresponding isotropic value is observed for $n_s$, which only deviates by $0.56$ standard deviations.

Note, however, that in this analysis we have not varied the new parameters that our model introduces, namely the value of the shear at the bounce $\sigma^2(t_B)$ and the total number of e-folds $N$ between the bounce and the onset of inflation (although $N$ could be fixed using the results of \cite{barrau}). We have rather fixed these two parameters in such a way that the quadrupolar modulation of our model agrees with the one observed  by Planck. A complete Bayesian analysis  should also include $\sigma^2(t_B)$ and  $N$ as free parameters, but such a calculation is  out of our current numerical capabilities, since MCMC methods  require to repeat thousands of times the calculations showed above to compute the angular correlation function, and each calculation takes about a week (see Sec. \ref{sec:numerics} for details of the computational cost). Nevertheless, our analysis suffices to show that the six cosmological parameters are largely insensitive to the new anisotropic features that the bounce introduces, when we restrict to configurations that respect the observational constraints.

\section{Discussion of the results and conclusions}\label{sec:discuss}

This paper introduces an extension of the standard cosmological model beyond general relativity, in which the big bang singularity is replaced by an {\em anisotropic} cosmic bounce. Our model is based on loop quantum cosmology, where the new physics producing the bounce originates from quantum gravitational effects. A complete loop quantization of Bianchi I spacetimes interacting with cosmic perturbations is out of reach at the present time. We have rather quantized cosmic perturbations propagating on the effective Bianchi I geometries of loop quantum cosmology. This strategy has been proven to capture accurately the physics of perturbations in FLRW spacetimes, and we have assumed that this is also the case for Bianchi I quantum geometries. We adopt a Fock quantization to deal with perturbations, neglecting potential polymer quantum effects that could affect the physics of perturbations in full quantum gravity. These are our main assumptions, and are a consequence of the lack of a complete theory of quantum gravity. However, we believe that they are physically reasonable, as long as perturbations remain small throughout the evolution. This is the case, as measured by the primordial power spectra $\mathcal{P}_{ss'}(\vec k)\ll 1$. Therefore, our treatment is well-aligned with the level of mathematical rigor that is common in studies of the early universe, and suffices to capture the main physical aspects of perturbations propagating across an anisotropic bounce.

The goal of the bounce in our model is not to replace inflation, but rather to complement it by removing the big bang singularity (a similar analysis without an inflationary phase, and where perturbations are generated before the bounce (see e.g.\ \cite{Khoury:2001wf,Lehners:2008vx,Brandenberger:2012zb,Raveendran:2017vfx,Ijjas:2016vtq}) can also be analyzed within the formalism presented here; it will be the focus of future work). In this model, the universe isotropizes both in the far past and future, but anisotropies dominate the bounce---we have restricted  to spacetimes compatible with a sufficiently long phase of inflation in the expanding  branch. We find that whenever anisotropies are large, gauge invariant cosmological perturbations are all coupled among themselves, and there is no universal way to disentangle scalar from tensor perturbations. However,  because anisotropies fall-off to the past and future of the bounce, perturbations decouple at early and late times. Furthermore, perturbations find an adiabatic regime in the past, before the bounce, for all the wavelengths that we can probe in the CMB. This implies that there is a preferred initial quantum state for these perturbations, which define the ``in'' Fock space of scalar and tensor perturbations, as described in detail in Sec. \ref{thevacuum}. Towards the end of the inflationary era in the expanding branch,  the universe is isotropic again, and one finds again a well defined ``out'' Fock space, built from the standard Bunch-Davies vacuum during inflation. Hence, this model provides a neat scenario where one has a well-defined notion of initial and final Fock spaces, and where one can compute the transition amplitudes of the ``in'' vacuum to different states in the ``out'' Fock space, i.e.\ the $\cal S$-matrix. The coupling between perturbations in the anisotropic phase generates quantum entanglement between scalar and tensor perturbations in the final state, as well as anisotropic features. We have solved the evolution of the system in detail, and showed that, although anisotropies in the spacetime are large only for a short interval around the bounce, perturbations retain memory of this anisotropic phase of the universe, preserving these features until the end of inflation and beyond. We have worked out the details of these features, and reported them in the form of angular correlation functions in the CMB, $C_{\ell \ell',mm'}^{XX'}$, where $X,X'=T,E,B$ denote temperature and the two components of the polarization of CMB photons. We have shown that the entanglement induced by anisotropies is manifest in cross-correlation functions for $X\neq X'$, while anisotropies also produce correlations with $\ell\neq \ell'$, that would vanish  in the isotropic limit. In particular, our model predicts nonzero correlations between temperature and  B-mode polarization (T-B), something that is forbidden by isotropy combined with parity  (our model respects parity, but not isotropy). 

We have contrasted our findings with current data from the CMB. The strongest constraints come from Planck's observations of a quadrupolar modulation \cite{plnck2018}. The observed quadrupole is in tension with isotropy, although the  significance is small and compatible with a statistical fluke within an isotropic universe. We have taken  the viewpoint that this quadrupole is a real feature in the CMB, which originates in primordial anisotropies, and have shown that our model is able to account for it and, furthermore, we  predict that its amplitude scales  as $1/k$ with the norm of the comoving wave number $\vec k$ of perturbations. Interestingly, the agreement of our model with the observed quadrupole  takes place when the number of $e$-folds between the bounce and the end of inflation coincides with results previously anticipated in \cite{barrau}. On the other hand, we find that the observed quadrupole imposes severe constraints on other anisotropic features in the CMB. We have computed all angular correlation functions $C_{\ell \ell',mm'}^{XX'}$ with $X,X'=T,E,B$, that our model predicts simultaneously with the observed quadrupole, and have discussed their  magnitude compared with the predictions of an isotropic universe. We have found that, although some of these anisotropic correlations and cross-correlations are small and difficult to be observed, others are not, and could be within the range of observations of future missions dedicated to measuring with precision the polarization of the CMB, such as CORE \cite{core}. The confirmation of some of the predictions we make here would increase the statistical  significance of the quadrupolar modulation observed by Planck, and  confirm that it is a relic of an anisotropic phase of the pre-inflationary universe. On the other hand, our model can also account in a natural manner for the quadrupolar asymmetry recently observed in the distribution of spin directions of galaxies \cite{shamir}. If these observations are confirmed, they will provide a strong motivation for primordial anisotropies.

Although we have used the spacetime predicted by the effective theory of loop quantum cosmology, we expect our results  to apply also to other bouncing  scenarios,  such as those explored in \cite{Mukhanov,llnwe, Shtanov:2002mb}. This will be the case as long as anisotropic contributions to the spacetime geometry  fall off sufficiently fast away from the bounce, and also if other finer details of the spacetime geometry around the bounce have subleading effects. The analysis presented here also adds robustness to previous studies related to the phenomenology of LQC based on isotropy, but it goes a step beyond by including for the first time anisotropic perturbations. Notice that even if anisotropic features in the CMB are eventually ruled out by future observations, our model can be used to explain in quantitative terms the length  of the inflationary phase needed to wash away all primordial anisotropies. 

To finish, the tools introduced in this manuscript, and in the companion paper  \cite{aos}, can be applied to study similar features in other models of the early universe. To further facilitate the application of our tools, we have made publicly available a code based on {\tt Mathematica} to derive gauge invariant perturbations and the equations of motion they satisfy in Bianchi I spacetimes \cite{ntbk} (see \cite{aos3} for a pedagogical description of this code), and  a second code based on the C programming language to numerically compute the evolution of perturbations, and to evaluate observable quantities in the CMB  \cite{num-lib}.\\
 
\acknowledgments
We have benefited from  discussions with Abhay Ashtekar,  Mar Bastero-Gil, Brajesh Gupt, Guillermo A. Mena Marug\'an, Jorge Pullin, Parampreet Singh and Edward Wilson-Ewing. This work is supported by the NSF CAREER grant No. PHY-1552603, Project. No. FIS2017-86497-C2-2-P of MICINN from Spain, and from funds of the Hearne Institute for Theoretical Physics.  V.S. was supported by Louisiana State University and Inter-University Centre for Astronomy and Astrophysics during earlier stages of this work. J.O. acknowledges the Operative Program FEDER 2014-2020 and the Consejer\'ia de Econom\'ia y Conocimiento of the Junta de Andaluc\'ia. Portions of this research were conducted with high performance computing resources provided by Louisiana State University (http://www.hpc.lsu.edu). 

\appendix

\section{Effective dynamics in LQC: some useful expressions}\label{app:lqc}

In this appendix we specify several phase space functions that are required for the evolution of perturbations on the effective geometries  of LQC. The background variables can be determined at any time  by solving  the effective equations, given in Eq. \eqref{effeqs}. From  them, one can easily verify that the directional Hubble parameters are given by
\begin{align}
H_1:=\frac{1}{2 \gamma 
   \sqrt{\Delta} } \left[\sin \left(\bar\mu_1c_1 -\bar\mu_2c_2 \right)+\sin \left(\bar\mu_1c_1 -\bar\mu_3c_3
   \right)+\sin \left(\bar\mu_2c_2 +\bar\mu_3c_3 \right)\right],
\end{align}

\begin{align}
H_2:=\frac{1}{2 \gamma 
   \sqrt{\Delta} } \left[\sin \left(\bar\mu_2c_2-\bar\mu_1c_1 \right)+\sin \left(\bar\mu_2c_2 -\bar\mu_3c_3
   \right)+\sin \left(\bar\mu_1c_1 +\bar\mu_3c_3 \right)\right],
\end{align}

\begin{align}
H_3:=\frac{1}{2 \gamma 
   \sqrt{\Delta}} \left[\sin \left(\bar\mu_3c_3-\bar\mu_1c_1
   \right)+\sin \left(\bar\mu_3c_3-\bar\mu_2c_2 \right)+\sin \left(\bar\mu_1c_1 +\bar\mu_2c_2 \right)\right].
\end{align}
From these expressions we obtain  the mean Hubble rate as $H=\frac{1}{3}\sum_{i=1}^3H_i$. The potentials that appear in Eq. \eqref{eqginper} are given by
\bea
\, {\cal U}_{00}&=&a^2\, V_{\phi\phi}\, - \frac{2 \kappa \, \pp^2 {\cal F}_2}{a^3}  + 2 \kappa  \, {\cal F}_1  \left(-\frac{\kappa\,\pp^2\,p_a}{3a^{5}}\, + \,2 \,V_{\phi}\, \pp\right),\\ \nonumber
   {\cal U}_{01}&=&{\cal U}_{10} \,= \,\frac{2\sqrt{\kappa}}{a^2}\left(-a^2\,\pp\, \sigma_{(5)} \, {\cal F}_2  + a^5V_{\phi} \, \sigma_{(5)}\, {\cal F}_1 - a^2\,\pp \, {\cal G}_{5}\,  {\cal F}_1\, +\, \frac{\kappa}{6}\,\pp\,p_a\,\sigma_{(5)}\,{\cal F}_1\right)\,,\\ \nonumber
   {\cal U}_{02}&=&{\cal U}_{20}\,= \,\frac{2\sqrt{\kappa}}{a^2}\left(-a^2\,\pp\, \sigma_{(6)} \, {\cal F}_2  + a^5\,V_{\phi} \, \sigma_{(6)}\, {\cal F}_1 - a^2\,\pp \, {\cal G}_{6}\,  {\cal F}_1\, +\, \frac{\kappa}{6}\,\pp\,p_a\,\sigma_{(6)}\,{\cal F}_1\right)\,,\\ \nonumber
   {\cal U}_{12}&=&{\cal U}_{21}\,=\, 2\,\sigma_{(5)}\,\sigma_{(6)}\,\left( a^2 -\, a^3\,{\cal F}_2\,+\, \frac{2}{3}\,\kappa\,a\,p_a\,{\cal F}_1\right) - \left( \, 2\,a^3\,\sigma_{(6)}\,{\cal G}_{5}\, +\, 2\,a^3\,\sigma_{(5)}\,{\cal G}_{6}\right)\,{\cal F}_1\, , \\ \nonumber
{\cal U}_{11}\, &=&\,- 2 \,a^2\, \sigma_{(6)}^2\, +\,\frac{\kappa p_a \,\sigma_{(2)}}{\sqrt{6}}\,-\,a^2\,\sqrt{\f{2}{3}}{\cal G}_2\, +\,\frac{4}{3}\,\kappa\,a\,p_a\,\sigma_{(5)}^2\,{\cal F}_1\,-\,4 \,a^3\, \sigma_{(5)}\,{\cal F}_1 \, {\cal G}_5\,-\,2\,a^3\, \sigma_{(5)}^2\,{\cal F}_2\, ,\\ \nonumber
{\cal U}_{22}\, &=&\,- 2 \,a^2\, \sigma_{(5)}^2\, +\,\frac{\kappa p_a \,\sigma_{(2)}}{\sqrt{6}}\,-\,a^2\,\sqrt{\f{2}{3}}{\cal G}_2\, +\,\frac{4}{3}\,\kappa\,a\,p_a\,\sigma_{(6)}^2\,{\cal F}_1\,-\,4 \,a^3 \,\sigma_{(6)}\, {\cal F}_1\, {\cal G}_6\,-\,2\,a^3\, \sigma_{(6)}^2\, {\cal F}_2\,, \ea
 where $V_{\phi}\equiv dV/d\phi$, $V_{\phi\phi}\equiv d^2V/d\phi^2$, and
\begin{eqnarray}
    {\cal F}_1\, &=&\, \frac{-\frac{\kappa p_a}{2a^3}\, +\,\sqrt{\frac{3}{2}} \,\frac{\sigma_{(2)}}{a}}{
      2\kappa\rho\,+\,\sigma_{(3)}^2\,+\, \sigma_{(4)}^2+\, \sigma_{(5)}^2\,+\, \sigma_{(6)}^2},\\ \nonumber
  {\cal F}_2\, &=&\frac{\frac{3\kappa \, V}{a}\, - \,\frac{\kappa^2 p_a^2}{3a^5}\,+\,\frac{\kappa p_a\sigma_{(2)}}{2\sqrt{6}a^3}\, +\, \sqrt{\frac{3}{2}}\frac{{\cal G}_{2}}{a}\,-\,{\cal F}_1\left[\frac{\kappa^2 \pp^2p_a}{a^8}\,+\,2\,\sigma_{(3)}\, {\cal G}_3\,+\,2\,\sigma_{(4)}\, {\cal G}_4+\, 2\,\sigma_{(5)}\, {\cal G}_5\,+\, 2\,\sigma_{(6)}\, {\cal G}_6)\right]}{
    2\kappa\rho\,+\,\sigma_{(3)}^2\,+\, \sigma_{(4)}^2+\, \sigma_{(5)}^2\,+\, \sigma_{(6)}^2},\\ \nonumber
   {\cal G}_2 &=& \frac{\kappa p_a\sigma_{(2)}}{2\,a^2}\, -\, \sqrt{\frac{3}{2}}\left(\sigma_{(3)}^2 \,+\, \sigma_{(4)}^2\right),\\ \nonumber
   {\cal G}_3 &=& \frac{\kappa\, p_a\, \sigma_{(3)}}{2\,a^2}\,  +\, \frac{1}{\sqrt{2}} \left(\sqrt{3}\sigma_{(2)} \sigma_{(3)} - \sigma_{(3)} \sigma_{(5)} - \sigma_{(4)} \sigma_{(6)}\right),\\ \nonumber
   {\cal G}_4 &=&  \frac{\kappa p_a\sigma_{(4)}}{2\,a^2} + \frac{1}{\sqrt{2}} \left(\sqrt{3}\sigma_{(2)} \sigma_{(4)} + \sigma_{(4)} \sigma_{(5)} - \sigma_{(3)} \sigma_{(6)}\right)\,, \\ \nonumber
   {\cal G}_5 &=& \frac{\kappa p_a\sigma_{(5)}}{2\,a^2}\, +\, \frac{1}{\sqrt{2}}(\sigma_{(3)}^2 - \sigma_{(4)}^2),\\ \nonumber
   {\cal G}_6 &=& \frac{\kappa p_a\sigma_{(6)}}{2\,a^2} + \sqrt{2}\,\sigma_{(3)}\sigma_{(4)}.
  \end{eqnarray}
These expressions are valid for both perturbations in classical GR and effective LQC. However, in the latter case, we must recall that the background phase space functions must be replaced by the ones in effective LQC. For instance, $p_a=-\frac{2}{\kappa}\, a^2 \, (H_1+H_2+H_3)$, as well as the components $\sigma_i$ of the shear tensor $\sigma_{ab}$,  are functions of the directional Hubble rates $H_i$, given in the expressions above. They appear implicitly in $\sigma_{(n)}:=\sigma_{ab}A_{(n)}^{ab}$ (with $n=1,\ldots,6$), the projections of the shear tensor $\sigma_{ab}$ on the matrices $A_{(n)}^{ab}$ obtained from
\begin{align}
 {A}^{{(1)}}_{ab}\, &=\, \f{h_{ab}}{\sqrt{3}}, \hspace{0.5in} & {A}^{(4)}_{ab}\, &=\,\f{1}{\sqrt{2}}\, \l(\, \hat{ k}_a\, \h y_b\, +\, \hat{ k}_b\, \h y_a \,\r),\nonumber\\
 {A}^{(2)}_{ab}\, &=\,\sqrt{\f{3}{2}}\,\l(\hat{ k}_a\,\hat{ k}_b - \f{h_{ab}}{3}\r), \hspace{.5in}  &{A}^{(5)}_{ab}\,& = \, \f{1}{\sqrt{2}}\, \l(\, \hat x_a\, \h x_b\, -\, \hat y_a\, \h y_b \,\r), \nonumber\\
{A}^{(3)}_{ab}\, &=\, \f{1}{\sqrt{2}}\, \l(\, \hat{ k}_a\, \h x_b\, +\, \hat{ k}_b\, \h x_a \,\r),
 \hspace{.5in} &{A}^{(6)}_{ab}\, &=\,\f{1}{\sqrt{2}}\, \l(\, \hat x_a\, \h y_b\, +\, \hat x_b\, \h y_a \,\r), \label{matrixbases}
\end{align}
by raising the spatial indices with $h_{ab}$ (the effective spatial metric of LQC). In addition, $\hat{k}$ is the unit vector in the direction of $\v{k}$, normalized with respect to $h_{ab}$. $\h x$ and  $\h y$ are two additional unit vectors that form, together with $\hat k$, a time-dependent orthonormal triad, with orientation given by $\hat x \times \hat y=\hat k$. For additional details, see Ref. \cite{aos}.

We would like to remark that the equations of motion of perturbations in classical GR are equivalent to the ones obtained in \cite{ppu-BI1}, modulo the background constraint. The ones adopted in our manuscript provide an evolution of perturbations that is well defined at all times, including the strong gravity regime in both GR (excluding the classical singularity) and effective LQC. 

\section{Mathematical aspects of spin-weighted spherical harmonics}\label{app:swsh}

The spin-weighted spherical harmonics $_{s} Y_{\ell m}$ are well known in the literature. They were introduced in Refs. \cite{swsh}, and we summarize here their main properties, mainly for convenience. They can be obtained from the standard spherical harmonics by
\bea
_{s} Y_{\ell m}&=&\left[\frac{(\ell-s) !}{(\ell+s) !}\right]^{\frac{1}{2}} \slashed{\partial}^{s} Y_{\ell m} \quad,(0 \leq s \leq \ell),\\
_{s} Y_{\ell m}&=&\left[\frac{(\ell+s) !}{(\ell-s) !}\right]^{\frac{1}{2}}(-1)^{s} \bar{\slashed{\partial}}^{-s} Y_{\ell m} \quad,(-\ell \leq s \leq 0),
\ea
where ${\slashed{\partial}}$ and $\bar{\slashed{\partial}}$ are the following differential operators
\bea
{\slashed{\partial}}&=&-\sin ^{s}(\theta)\left[\frac{\partial}{\partial \theta}+i \csc (\theta) \frac{\partial}{\partial \phi}\right] \sin ^{-s}(\theta),\\
\bar{\slashed{\partial}}&=&-\sin ^{-s}(\theta)\left[\frac{\partial}{\partial \theta}-i \csc (\theta) \frac{\partial}{\partial \phi}\right] \sin ^{s}(\theta).
\ea
They satisfy the following properties
\bea\label{eq:Y0}
{}_{s} \bar Y_{\ell m}&=&(-1)^{s+m} {}_{-s}Y_{\ell-m},\\
\slashed{\partial}\, {}_{s}Y_{\ell m}&=&[(\ell-s)(\ell+s+1)]^{\frac{1}{2}} {}_{s+1} Y_{\ell m},\\
\bar{\slashed{\partial}}\, {}_{s}Y_{\ell m}&=&-[(\ell+s)(\ell-s+1)]^{\frac{1}{2}} {}_{s-1} Y_{\ell m},
\ea
as well as the following relation under spatial inversion
\be\label{eq:sY-parity}
{}_{s} Y_{\ell m}(-\hat x)=(-1)^{\ell}{}_{-s}Y_{\ell m}(\hat x).
\ee
An explicit expression for them is
\bea\nonumber
{}_{s} Y_{\ell m}(\theta,\phi)&=&e^{i m \phi}\left[\frac{(\ell+m) !(\ell-m) !}{(\ell+s) !(\ell-s) !} \frac{2 \ell+1}{4 \pi}\right]^{1 / 2} \sin ^{2 \ell}(\theta / 2)\\
&&\times \sum_{r} \left( \begin{array}{c}{\ell-s} \\ {r}\end{array}\right) \left( \begin{array}{c}{\ell+s} \\ {r+s-m}\end{array}\right)(-1)^{\ell-r-s+m} \cot ^{2 r+s-m}(\theta / 2).
\ea

Finally,  we have made use of triple integrals of spin weighted spherical harmonics, in our calculations of the angular correlation functions. They can be written in terms of well-known 3-$j$ symbols as
\be\label{eq:3jsym}
\int d\Omega_{\h k}\,{}_{s''}Y_{\ell'' m''}(\hat{k}) {}_{s}Y_{\ell m}(\hat{k}){}_{s'}Y_{\ell' m'}(\hat{k}) = \sqrt{\frac{\left(2\ell''+1\right)\left(2\ell+1\right)\left(2\ell'+1\right)}{4\pi}}
\begin{pmatrix}
  \ell'' & \ell & \ell'\\
  m'' & m & m'
\end{pmatrix}
\begin{pmatrix}
  \ell'' & \ell & \ell'\\
  -s'' & -s & -s'
\end{pmatrix},
\ee
if $s+s'+s''=0$. Actually, it is not difficult to check from this expression the following orthogonality relations of two spin weighted spherical harmonics
\be
\int d\Omega_{\h k}\  {}_{s}\bar Y_{\ell m}(\hat{k}){}_{s}Y_{\ell' m'}(\hat{k})=
\delta_{\ell\ell'}\delta_{mm'}.
\ee



\end{document}